\documentclass[%
reprint,
superscriptaddress,
preprintnumbers,
 amsmath,
 amssymb,
 aps,
 prd,
]{revtex4-1}
\usepackage{slashed}
\usepackage{graphicx}% Include figure files
\usepackage{dcolumn}% Align table columns on decimal point
\usepackage{bm}% bold math
\usepackage{hyperref}% add hypertext capabilities
\usepackage[mathlines]{lineno}% Enable numbering of text and display math
\newcommand{\bfs}{\mathbf{s}_\perp}
\newcommand{\bfk}{\mathbf{k}_\perp}
\newcommand{\bfl}{\boldsymbol{\ell}_\perp}

\newcommand{\bfq}{\mathbf{q}_\perp}
\newcommand{\bfp}{\mathbf{q}_\perp}
\newcommand{\bfb}{\mathbf{b}_\perp}

\newcommand{\bfy}{\mathbf{y}_\perp}

\usepackage{xcolor}
\usepackage[normalem]{ulem}
\begin{document}
\graphicspath{{Publication/PubFigs/}{./}}
\preprint{LA-UR-23-24141}

\title{eHIJING: an Event Generator for Jet Tomography in Electron-Ion Collisions }

\author{Weiyao Ke}
\email{weiyaoke@ccnu.edu.cn}
\affiliation{Institute of Particle Physics and Key Laboratory of Quark and Lepton Physics (MOE), Central China Normal University, Wuhan, Hubei, 430079, China}
\affiliation{Theoretical Division, Los Alamos National Laboratory, \\Los Alamos NM 87545, United States}
\affiliation{Nuclear Science Division, MS 70R0309, Lawrence-Berkeley National Laboratory, \\1 Cyclotron Rd, Berkeley CA 94720, United States}
\affiliation{Physics Department, University of California, Berkeley, California 94720, USA}

\author{Yuan-Yuan Zhang}
\email{zhangyuanyuan@cuhk.edu.cn}
\affiliation{
School of Science and Engineering, The Chinese University of Hong Kong, Shenzhen 518172, China}
\affiliation{University of Science and Technology of China, Hefei 230026, China}

\author{Hongxi Xing}
\email{hxing@m.scnu.edu.cn}
\affiliation{Guangdong Provincial Key Laboratory of Nuclear Science, Institute of Quantum Matter, South China Normal University, Guangzhou 510006, China}
\affiliation{Guangdong-Hong Kong Joint Laboratory of Quantum Matter,
Southern Nuclear Science Computing Center, South China Normal University, Guangzhou 510006, China}

\author{Xin-Nian Wang}
\email{xnwang@lbl.gov}
\affiliation{Nuclear Science Division, MS 70R0309, Lawrence-Berkeley National Laboratory, \\1 Cyclotron Rd, Berkeley CA 94720, United States}
\affiliation{Physics Department, University of California, Berkeley, California 94720, USA}

\date{\today}

\begin{abstract}
We develop the first event generator, the electron-Heavy-Ion-Jet-INteraction-Generator (eHIJING), for the jet tomography study of electron-ion collisions. In this generator, energetic jet partons produced from the initial hard scattering undergo multiple collisions with the nuclear target. The collision rate is proportional to the transverse-momentum-dependent (TMD) gluon density in the nucleus, which is given by a simple model inspired by the physics of gluon saturation. Medium-modified QCD splitting functions within the higher-twist (HT) and generalized higher-twist (GHT) frameworks are utilized to simulate parton showering in the nuclear medium that takes into account the non-Abelian Landau-Pomeranchuck-Midgal interference effect. 
Employing eHIJING, we revisit hadron production in semi-inclusive deep inelastic scattering (SIDIS) as measured by EMC, HERMES, and recent CLAS experiments. 
eHIJING with both GT and GHT frameworks gives reasonably good descriptions of these experimental data. 
Predictions for experiments at the future electron-ion colliders are also provided. It is demonstrated that future measurements of the transverse momentum broadening of single hadron spectra can be used to map out the two-dimensional kinematic ($Q^2, x_B$) dependence of the jet transport coefficient $\hat{q}$ in cold nuclear matter.
\end{abstract}

\maketitle

\section{Introduction}

Understanding the parton dynamics in the nuclear matter is key to the programs at the future electron-ion collider \cite{Boer:2011fh,Accardi:2012qut}. Jet and hadron tomography in electron-nucleus collisions are of great importance in the study of nuclear partonic structures, jet transport coefficient, and the hadronization mechanism inside nuclear matter. Many progresses have already been made recently in this direction, including the extraction of the jet transport parameter in nuclei \cite{Wang:2002ri,Deng:2009ncl,Chang:2014fba,Ru:2019qvz,Ru:2023ars}, nuclear parton distributions functions (nPDF) \cite{Eskola:2009uj,Kovarik:2015cma,Eskola:2016oht,AbdulKhalek:2019mzd,Ethier:2020way} and fragmentation functions (nFF) \cite{deFlorian:2007aj,Sassot:2009sh}, and the nuclear transverse-momentum-dependent (TMD) parton distribution functions (TMD-PDF) and fragmentation functions (TMD-FF) \cite{Alrashed:2021csd} from global data analysis including $e+A$ fixed target experiments.

Some of these phenomenological studies with nuclear targets assume a factorization formula similar to that in the vacuum, then, the observed differences between $e$+$p$ and $e+A$ collisions are attributed to the nPDF and nFF. However, in such analysis, one should be careful of distinguishing intrinsic non-perturbative nuclear properties from dynamical nuclear modifications of jet/hadron production. These dynamical effects, originating from multiple jet-medium interactions of both partonic and hadronic nature, can be process-dependent. It is therefore essential to understand these contributions from both theoretical and modeling perspectives to improve the predictive power of the calculation, extract universal dynamical quantities of the cold nuclear matter, and eventually understand the intrinsic non-perturbative nature of nuclei in high-energy collisions.

In the field of relativistic heavy-ion collisions, parton propagation and jet modification in the hot and dense quark-gluon plasma have been the focus of both theoretical and experimental studies over several decades. For reviews on such topics, see Refs. ~\cite{Majumder:2010qh,Qin:2015srf,Blaizot:2015lma,Cao:2020wlm,Apolinario:2022vzg}. Multiple interactions between jet partons and the QGP medium lead to parton energy loss and the suppression of large transverse momentum single inclusive hadron and jet spectra, modified di-jet/di-hadron and $\gamma$-jet/hadron correlations, modification of jet fragmentation functions, jet shape and jet substructures. These observed phenomena have been predicted by theoretical models based on perturbative QCD (pQCD) calculations of parton transport through multiple scatterings. However, the most detailed test of our understanding of the medium-modified jet fragmentation function is only effective and subject to a large uncertainty. This is because one cannot precisely determine the initial jet energy in heavy-ion collisions even using the rare $\gamma/Z$-tagged jets, due to initial state radiation and complicated event activity. Furthermore, one must know the space-time evolution of the hot QGP medium, which is normally provided by hydrodynamic model simulations. Though these hydrodynamic models \cite{Luzum:2008cw, Shen:2014vra, Pang:2018zzo} are constrained by experimental data on soft bulk hadron spectra, uncertainties in the model parameters will also propagate to the calculation of jet modifications. In addition, contributions to the final jet energy by soft hadrons from the jet-induced medium response are non-negligible and, therefore should also be considered \cite{Cao:2020wlm}. 
Recent Monte Carlo models for the study of jet quenching in heavy-ion collisions~\cite{Li:2010ts,Zapp:2012ak,He:2015pra,Chen:2017zte,Putschke:2019yrg} are designed to take into account these effects.

\begin{figure}
    \centering
    \includegraphics[width=.9\columnwidth]{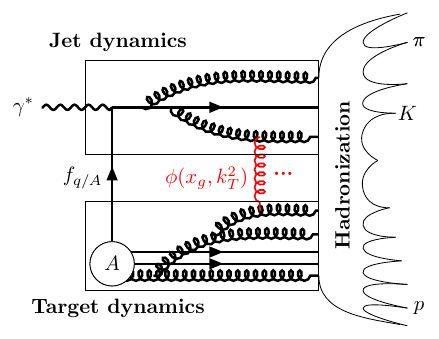}
    \caption{(Color online) A schematic plot for the hard process, jet evolution, target dynamics, Glauber gluon exchange, and the hadronization in a semi-inclusive DIS process.  eHIJING focuses on the jet evolution and does not include Target dynamics.}
    \label{fig:cartoon1}
\end{figure}

In the semi-inclusive DIS process, the initial jet energy can be determined from the scattered lepton. The cold nuclear medium probed by the energetic partons from the hard lepton-quark scattering is the ground state of an atomic nucleus.
Furthermore, in a collider experiment, the medium is also highly boosted, which separates the jet and target fragmentation in different phase-space regions.
Therefore, the semi-inclusive DIS (SIDIS) process can provide highly differential measurements of nuclear-modified jet fragmentation processes and powerful tests on various assumptions of parton-nuclear interactions.

Consider a quark jet produced at large Bjorken $x_B$ and hard scale $Q$ in SIDIS with a nuclear target, it acquires a large momentum in the nuclear rest frame $\nu=Q^2/(2x_B M_N)$. Multiple collisions between the large-momentum quark and the target are forward scatterings mediated by Glauber gluons. Glauber gluons are off-shell and carry a fraction $x_g$ of the nucleon's light-cone momentum $P_N^-$ that is much smaller than its transverse momentum $x_g P_N^-\ll \bfk$. 
The collision probability grows with the linear size of the nucleus $L\approx A^{1/3}\times 1.2~\textrm{fm}$, leading to jet/hadron momentum broadening $\Delta \langle \bfp^2\rangle_{eA}\propto A^{1/3}$ as observed in experiments \cite{HERMES:2007plz,HERMES:2009uge}. 
The momentum broadening can be related to the nuclear transverse-momentum dependent gluon distribution function $G(x_g, \bfk)$ at small $x_g$ \cite{Collins:1981uw, Ji:2002aa,Belitsky:2002sm,Casalderrey-Solana:2007xns,Liang:2008vz}. 

Multiple collisions will also modify the development of the parton shower. The key theoretical inputs are the medium-modified parton splitting functions induced by multiple collisions, which can be calculated in pQCD \cite{Guiot:2020vsf,Zhang:2003wk,Deng:2009ncl,Chang:2014fba,Li:2020rqj,Li:2020zbk} within the higher-twist framework \cite{Guo:2000nz,Wang:2001ifa,Wang:2002ri},  soft-collinear effective theory with Glauber gluons \cite{Idilbi:2008vm,Ovanesyan:2011xy}, as well as the most recent generalized higher-twist approach \cite{Zhang:2021kym,Zhang:2021tcc}.
The modified parton shower approach has provided a good quantitative understanding of the observed nuclear modification of the final fragmentation functions in SIDIS off nuclear targets \cite{HERMES:2000ytc, HERMES:2003icw, HERMES:2005mar, HERMES:2007plz}.
It may seem surprising that the problem can be treated in a perturbative manner, considering the average momentum broadening of a hadron in a nucleus is only a few hundred MeV. Such concern has been addressed in a recent study \cite{Ke:2023ixa} where it is found that, with a large enough $\nu$, $\sqrt{\nu/L}$ emerges as a semi-hard scale of the medium-induced parton splitting. This provides the foundation of a perturbative treatment of the medium-modified parton shower.

Eventually, the parton shower undergoes hadronization. The formation time of a light hadron carrying energy fraction $z_h$ of the parton is on the order $\tau_h \sim z_h\nu/\Lambda_{\rm QCD}^2$. If $\nu$ is large such that for most hadrons $\tau_h\gg L$, then to leading power of $L/\tau_h$, one can make the approximation that the hadronization process happens outside the nuclear medium and is still dominated by the fragmentation mechanism in the vacuum. In the other limits $\tau_h<L$, hadrons will form inside the nucleus and one has to consider hadron-level final-state interactions. Non-perturbative dynamical models, such as the hadronic transport approach with time-dependent pre-hadron cross-section~\cite{Gallmeister:2007an,Buss:2011mx} have been developed for this purpose.

In this work, we will focus on SIDIS in the large $\nu$ region and develop the eHIJING (electron-Heavy-Ion-Jet-INteraction-Generator) event generator for simulations of jet production in $e+A$ collisions. 
Fig.~\ref{fig:cartoon1} is a schematic plot showing the physics included in eHIJING:
\begin{itemize}
\item The nuclear collinear or transverse momentum integrated PDFs will be given by the EPPS parameterization~\cite{Eskola:2009uj,Eskola:2016oht}.
\item The distribution of Glauber gluons that collide with jets is modeled by a TMD gluon distribution in the small $x$ region \cite{Collins:1981uw,Ji:2002aa,Belitsky:2002sm} as motivated by the gluon saturation model \cite{Mueller:1989st,Golec-Biernat:1998zce}. Note that this model does not include the dynamical evolution on the target side.
\item  Jet evolution is simulated within both the higher-twist (HT) \cite{Guo:2000nz,Wang:2001ifa,Zhang:2003wk,Majumder:2009ge} and generalized high-twist (GHT) \cite{Zhang:2021kym,Zhang:2021tcc} framework.
\item At the moment, eHIJING does not include any hadron-level final-state interactions. This can be pursued in the future for an improved description of SIDIS in the lower $\nu$ region.
\end{itemize}

We will apply eHIJING to study the medium modifications of unpolarized SIDIS measurement at CLAS, HERMES, and EMC experiments which generally involve DIS at large $x_B$. Furthermore, we test different assumptions and approximations in the simulation of modified jet evolution in the medium. This allows us to estimate the theoretical uncertainty of jet tomography studies. Future high-precision determination of TMD observables at the Electron-Ion Collider (EIC) can provide better constraints on these calculations that will, in turn, improve the theoretical accuracy of jet tomography in $A$-$A$ collisions. 

The remainder of this paper is organized as follows. Section \ref{sec:overview} gives an overview of the physical ingredients and design of eHIJING. Section \ref{sec:model:coll} thoroughly describes the relation between multiple collisions and the TMD gluon distribution at small-$x$ and the stochastic implementation in eHIJING. Two types of in-medium QCD splitting functions in the generalized higher-twist and higher-twist framework are described in Section \ref{sec:model:rad}.
In Section \ref{sec:model:evo}, we demonstrate the implementation of the modified QCD splitting functions in the jet parton shower and fragmentation. In Section \ref{sec:results} we present and discuss the results from eHIJING simulations as compared to available data from the EMC, HERMES, and CLAS experiments. We make projections for future experiments at EIC and EicC (EIC in China) and discuss future improvements in Section \ref{sec:eic}. Summaries are given in Section \ref{sec:summary}.

\section{Overview of the eHIJING event generator}
\label{sec:overview}
\begin{figure}
    \centering
    \includegraphics[width=\columnwidth]{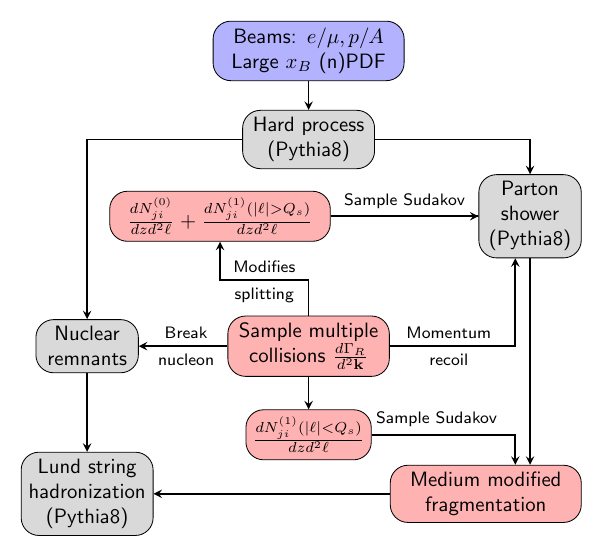}
    \caption{(Color online) Flow chart of eHIJING Monte Carlo model. Blocks shown in grey are ingredients needed$e$+$p$ simulations in Pythia8. Blocks in red are eHIJING's implementation of medium effects. }
    \label{fig:flowchart}
\end{figure}

In Fig.~\ref{fig:flowchart}, we outline the eHIJING simulation by a flow chart. If one omits the blocks colored in red, the rest of the flowchart represents the generation of an $e$+$p$ event.
In eHIJING, the $e$+$p$ collision is handled by the Pythia8235 event generator \cite{Sjostrand:2014zea,Cabouat:2017rzi}. It includes the generation of the hard process, the development of the vacuum parton shower, the handling of the hadronic remnant, and hadronization using the Lund string model.

The triggering event in the LO perturbative parton model is the ``knock out'' of a quark. Given the four-momenta of the incoming and the outgoing leptons, $\ell_1$ and $\ell_2$, respectively, and the momentum per nucleon $p$ of the nucleus with the atomic number $A$, the LO DIS cross section is
\begin{align}
E_{\ell_2}\frac{d\sigma_{\rm DIS}^{ep}}{d^3\ell_2}
&=\frac{\alpha^2_{\rm EM}}{2\pi s}\frac{4\pi}{Q^4} \sum_q e_q^2 f_{q/p}(x_B,Q^2) \nonumber \\
&\times L^{\mu\nu} (x_B e^L_{\mu\nu} - \frac{1}{2}e^T_{\mu\nu})\; ,
\label{sigma-dis}
\end{align}
where $\alpha_{\rm EM}$ is the electromagnetic fine-structure constant, $Q^2=-q^2$ with $q$ the momentum of the virtual photon $q=\ell_2-\ell_1$, $s=(p+\ell_1)^2$ is the total invariant mass of the lepton-nucleon system.
\begin{align}
  e^T_{\mu\nu}& = g_{\mu\nu}-\frac{q_\mu q_\nu}{q^2}, \nonumber \\
  e^L_{\mu\nu}& = \frac{1}{p\cdot q} 
\left(p_\mu-\frac{p\cdot q}{q^2} q_\mu\right)
\left(p_\nu-\frac{p\cdot q}{q^2} q_\nu\right) \, .
\label{eq-tensor}
\end{align}
The leptonic tensor $L_{\mu\nu}$ is given by  
\begin{align}
L_{\mu\nu}=\frac{1}{2}\, {\rm Tr}(\slashed{\ell}_1 \gamma_{\mu}
\slashed{\ell}_2 \gamma_{\nu}) \; .
\end{align}
The Bjorken variable $x_B$ is defined as $x_B=Q^2/2p\cdot q$ and
 $f_{q/p}(x_B, Q^2)$ is the collinear PDF of quark at scale $Q^2$. Integrating over the angle of the out-going electron, the DIS cross section at leading order can be expressed as
 \begin{align}
     \frac{d\sigma_{\rm DIS}^{ep}}{dQ^2dx_B}&=\frac{4\pi\alpha^2_{\rm EM}}{Q^4} \sum_q e_q^2 f_{q/p}(x_B,Q^2) \nonumber \\
     &\times \left(1-y+\frac{y^2}{2}-\frac{m_N^2y^2}{Q^2}\right),
 \end{align}
 where $m_N$ is the nucleon mass and $y=p\cdot q/p\cdot \ell_1=Q^2/[(s-m_N^2)x_B]$ is usually referred to as the inelasticity of the collision.
The inclusive DIS cross-section for $e$+$A$ collision is obtained by replacing $f_{q/p}$ with the collinear nuclear PDF $f_{q/A}$.

\begin{figure}
\centering
\includegraphics[scale=.85]{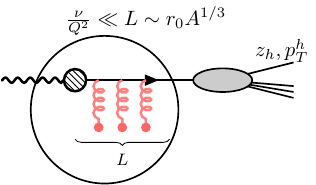}\hfill\includegraphics[scale=.85]{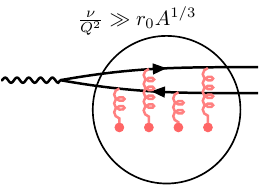}
\caption{(Color online) The space-time picture of a quark knock-out process at large $x_B$ (left)  and a di-jet production process at small $x_B$ (right). Red vertical coils indicate multiple collisions between the jet parton and the nucleus.}
\label{fig:spacetime}
\end{figure}

The semi-inclusive DIS process measures a hadron or a jet in the hadronic final state, in addition to measuring the deflected lepton. To take the single-hadron production as an example, the factorization formula for the $z_h$-differential cross-section at leading order is
\begin{align}   
&\frac{d\sigma_{\rm DIS}^{e+A\rightarrow h}}{dx_B dQ^2 dz_h} = \frac{4\pi\alpha_{\rm EM}^2 }{Q^4}  \sum_{q} e_q^2 f_{q/A}(x_B, Q^2) \nonumber\\
&\times   \left(1-y+\frac{y^2}{2}-\frac{m_N^2y^2}{Q^2}\right) d_{h/q}(z_h, Q^2).
\end{align}
$z_h=E_h/\nu$ is the fraction of photon energy carried by the hadron. $d_{h/q}(z_h, Q^2)$ is the collinear fragmentation function. 
For the $z_h$ and $p_T^h$ differential production of hadron (as illustrated on the left of Fig. \ref{fig:spacetime}), one should refer to the TMD factorization formula for the SIDIS process, e.g., see Ref. \cite{Boussarie:2023izj}.

By comparison, in Pythia8, the LO cross-section is generated first. The QCD evolution, corresponding to the scale evolution of $f_{q/A}$ and $d_{h/q}$, is treated in the transverse-momentum-ordered parton shower approach. It uses the QCD splitting function to recursively generate parton branching from the hard scale down to a cut-off scale $Q_0\gtrsim \Lambda_{\rm QCD}$.
Finally, the Lund string fragmentation model handles the hadronization of the color-neutral system that includes both the parton shower and the beam remnants. 
As for transverse-momentum-dependent observables, the event generator models 1) recoils from perturbative parton branching, 2) non-perturbative transverse momentum of hadron production from the Lund-string model, and 3) a non-perturbative model that gives the initial-state quark a primordial transverse momentum inside the nucleon \cite{Sjostrand:2004pf}.

The red blocks in Fig. \ref{fig:flowchart} represent eHIJING's modification to the event generator for $e$+$A$. At the center of the modification is a model for sampling the multiple collisions between the jet parton and the nucleus, and it will be explained in detail in Sec. \ref{sec:model:coll}. 
The multiple collisions further modify the splitting function (see Sec. \ref{sec:model:rad}).
How the modified splitting functions affect the parton shower development at both high and low virtualities is explained in Sec. \ref{sec:model:evo}. The nucleons' remanent from the multiple collisions and hadronization are discussed in Sec. \ref{sec:model:remnant}.

For event generation in $e$+$A$ collisions, this work will focus on the kinematic region with large $x_B$ and high $Q^2$ while keeping $\nu$ large.
This ensures that the hard production process is localized in the nucleus, i.e.,
\begin{align}
&\Delta r_{\perp} \sim \frac{1}{Q} \ll L, \\
&\Delta r^+ \sim \frac{\nu}{Q^2} = \frac{1}{2x_B m_p} \approx \frac{0.1}{x_B} \textrm{~fm} \ll L,\label{condition-2}
\end{align}
where $L$ is the typical path length that the quark propagates in the nucleus.
For a spherical heavy nucleus $A$, the average path length $\langle L \rangle = 3/4 r_0 A^{1/3}$ with $r_0\approx 1.2$ fm. Therefore, the second inequality is satisfied for $x_B\gg 0.1/A^{1/3}$.
Furthermore, the hadron formation time is long compared to the path length
\begin{align}
\tau_h \sim \frac{z_h\nu}{\Lambda_{\rm QCD}^2} \gg L,
\end{align}
so the hadronization mechanism is dominated by fragmentation in the vacuum.
This is the perfect region to study the effect of parton transport in nuclear matter, as shown on the left of Fig.~\ref{fig:spacetime}.

At smaller $x_B$, the hard process can be coherent over several nucleons, and one needs to include the nuclear shadowing effect, e.g., included by the empirical nuclear PDF \cite{Eskola:2016oht} or from resumed power correction calculations \cite{Qiu:2003vd}.
In addition, the di-jet production from NLO processes becomes important. Eventually, for $x_B \ll 0.1/A^{1/3}$, the interaction is dominated by the dipole reaction: virtual photon fluctuates into a $q\bar{q}$ pair and interacts with the whole nucleus coherently, as shown on the right of Fig.~\ref{fig:spacetime}. The average path length is $3/2 R_A$, twice the average path length for DIS at large $x_B$.
This regime is beyond the scope of the current work of eHIJING, but there are other specialized event generators developed for small-$x_B$ physics, for example, see Ref. \cite{Shi:2022hee}. 
For moderate $x_B$, it will be interesting to investigate how to interpolate the two different space-time pictures of DIS in the future.

From the space-time picture illustrated on the left of figure \ref{fig:spacetime}. We included the following nuclear effects in eHIJING:
\paragraph{Nuclear PDFs.} Nucleon motion and correlations inside a nucleus can alter the effective quark distribution function per nucleon at large $x_B$ \cite{Szczurek:2003yv,Seely:2009gt} and coherent scatterings can lead to nuclear shadowing effect \cite{Qiu:2003vd}. We include these effects by using the parameterized nuclear PDFs that include the effect of Fermi motion, EMC, and (anti-) shadowing~\cite{Eskola:2009uj,Kovarik:2015cma,Eskola:2016oht}. 
Some dynamical models can systematically describe the nuclear shadowing \cite{Qiu:2003vd,Qiu:2004qk,Qiu:2004da}, which can be considered as alternative models in the future.
Of course, these are mostly effects at the level of single-parton/nucleon distribution function, we are still missing correlation information of the nucleus, such as short-range nucleon correlation \cite{CiofidegliAtti:2015lcu,Hen:2016kwk}.

\paragraph{Multiple collisions.} When an energetic jet propagates through the nuclear medium, partons in the jet shower will encounter multiple collisions with the nuclear target remnants. The corresponding collision rate for parton $a$ is related to the transverse-momentum-dependent (TMD) gluon distribution density at small $x$
\begin{align}
\frac{d\Gamma_a}{d^2\bfk} = \frac{C_2(a)}{d_A} \rho_N \alpha_s \frac{\phi_g(x_g, \bfk)}{\bfk^2},
\label{eq:multiple_collisions}
\end{align}
where $\rho_N$ is the nucleon density inside the nucleus with atomic number $A$, $\phi_g$ is the effective TMD gluon distribution function per nucleon, $d_A=N_c^2-1$ and $C_2(a)$ is the quadratic Casimir in the color representation of parton $a$. For a quark $C_2(q)=C_F=(N_c^2-1)/2N_c$ and $C_2(g)=C_A=N_c$ for a gluon.  The momentum fraction carried by these exchanged gluons $x_g = x_B\bfk^2/Q^2$ is small, and gluon number density can be large. The emergent gluon saturation scale $Q_s^2 \propto A^{1/3}$ dictates the typical scale of $\bfk$ in this regime \cite{Mueller:1999wm}. When $Q_s^2\gg \Lambda_{\rm QCD}$, a weakly-coupled model calculation of $\phi_g(x_g, \bfk)$   is possible \cite{McLerran:1993ni,McLerran:1994vd}. Therefore, we will use a saturation-motivated ansatz to model $\phi_g(x_g, \bfk)$ and generate multiple collisions for propagating shower partons. When each new parton is created in the hard process or the parton shower, a sequence of multiple collisions is sampled based on Eq.~(\ref{eq:multiple_collisions}).

\paragraph{Modified parton shower and fragmentation.}  Multiple collisions will modify the QCD splitting functions in the medium. We will use the higher-twist and generalized higher-twist results for the medium modified splitting function $\Delta P_{ij}$, which will be implemented into eHIJING by modifying the $\bfl$-ordered parton shower and fragmentation in Pythia8. For this, we adopt a similar idea from Ref.~\cite{Chang:2014fba} to model the in-medium collinear fragmentation function. Medium-modified splittings with a transverse momentum  $|\bfl|$ larger than $Q_s$ will be added to the Pythia8 parton shower program, while the modifications with transverse momentum smaller than $Q_s$ are handled by a separate routine after the parton level Pythia8 simulation is finished, leading to modified parton fragmentation. With this implementation of the modified jet shower, we can study the medium-modified transverse-momentum-dependent fragmentation. 

\paragraph{Nuclear excitation} Multiple collisions also excite the nuclear target. We assume that nucleons that participate in multiple collisions will be broken into recoiled pairs of quark and di-quark. They carry the respective color charge of the exchanged gluon to maintain the color neutrality of the entire system. 
The subsequent dynamics of the nuclear target are not considered in eHIJING.
For the physics of target dynamics, one may refer to recent studies with the BeAGLE event generator \cite{Zheng:2014cha,Morozov:2018voq,Chang:2022hkt,Li:2023dhb}.
Besides the nuclear dynamics, we have also omitted hadronic interactions between the jet and the nucleus. They can be important in 1) collisions with lower beam energy, where a significant fraction of hadrons forms inside the nucleus, 2) heavy flavor production in which heavy quarks travel at non-relativistic speed in the medium and hadronize before they exit the nuclear medium. One can couple the current eHIJING with a hadronic transport model in the future to study related physics.

Finally, there are some subtle issues when we use the DIS mode of Pythia8 in eHIJING, and we have changed a few Pyhtia8 default DIS settings:
\begin{itemize}
\item In Pythia8, a method called the ``dipole recoil'' is used to handle the four-momentum conservation in parton branching in DIS. Compared to the ``global recoil'' mode often used for initial-state radiations in hadronic collisions, it can reproduce the singular structure of the NLO DIS matrix-element calculations \cite{Cabouat:2017rzi}. 
In the ``dipole recoil'' approach, only the initial-state quark is taken as the emitter in parton branching, while the momentum of the final-state quark (the recoiler) will be adjusted to restore energy-momentum conservation. However, the medium interactions only affect the final-state quark so it is natural to choose the final-state quark as the radiator in the medium.
Therefore, we have chosen to use the global recoil mode of Pythia8 in eHIJING, even though the dipole recoil option is the recommended default choice for DIS.
There are some known problems with the global recoil, such as the uncertainty in $Q^2$ determination and the matrix-element matching. For this reason, we include App. \ref{sec:default-vs-dipole} and assess the difference between the two recoil options.
\item With both initial and final-state radiation switched on, Pythia8 by default interleaves initial and final-state radiation \cite{Sjostrand:2004ef}, where the transverse momentum of splitting orders initial and final-state emissions. To include medium corrections, it is more natural to treat the medium-modified final-state emission with the final-state multiple interactions after the initial-state radiations. Therefore, the ``interleaving'' option is turned off.
\item Other changes involve default fragmentation parameters, which we will elaborate in Sec. \ref{sec:results:baseline}.
\end{itemize}

\section{Multiple collisions and nuclear TMD gluon distribution}
\label{sec:model:coll}

In the Breit frame, the nucleus consists of highly boosted but transversely localized nucleons with weak correlation, as described by a one-nucleon density distribution $\rho(r^+, \mathbf{r})$.
The four momenta of the nucleon $p$, virtual photon $q$, quark $l_q'$ before and $l_q$ after the collisions are
\begin{align}
p&=\left[0,~\frac{Q^2}{2q^+ x_B},~\mathbf{0}\right],\label{eq:jet-side-p}\\
q &= \left[q^+,~-\frac{Q^2}{2q^+},~\mathbf{0}\right],\\
l_q' &= x_Bp=\left[0,~ \frac{Q^2}{2q^+},~ \mathbf{0}\right],\\
l_q &= q+\l_q' = \left[q^+,~ 0,~\mathbf{0}\right].
\end{align}
In this paper, we choose the convention for the product of two light-cone four vectors $[a^+,a^-, \mathbf{a}_\perp]$ and $[b^+,b^-, \mathbf{b}_\perp]$ as 
\begin{align}
a\cdot b=a^+b^-+a^-b^+ -\mathbf{a}_\perp\cdot\mathbf{b}_\perp.
\end{align}
The medium gluon that interacts with the outgoing quark has four-momentum
\begin{align}
k = [k^+=0, k^-, \bfk].
\label{eq:jet-side-k}
\end{align}
$(l_q+k)^2=0$ determines 
\begin{align}
k^- = \frac{\bfk^2}{2q^+} = \frac{\bfk^2}{Q^2}x_B p^- \equiv x_g p^-
\end{align}
$x_g= x_B \bfk^2/Q^2$ is the gluon's light-cone momentum fraction. 
In the interaction with collinear jet partons that are highly boosted in the plus direction, $k^+$ is power suppressed compared to the $p^+$ of the jet parton and is consistently set to zero. 
Nevertheless, a small but finite $k^+$ can be important in events with small jet energy. We will consider its impact during qualitative analyses in Sec.~\ref{sec:results:eA-dpT2}.

In a large nucleus, we only consider nuclear effects that are enhanced by the nuclear size $A^{1/3}$. Therefore, we neglect correlations from interactions between quark and gluon fields within the same nucleon. Then, one can effectively factorize the amplitude for quark production into the hard part and the additional quark-gluon rescattering. 
Consider a quark produced at coordinate $r_0=(r_0^+,\bfb$). $\bfb$ is the impact parameter as shown in Fig. \ref{fig:cartoon1}.
It rescatters with another nucleon at the location $r=(r^+, \bfb)$ exchanging momentum $k$ and the final momentum $l_q$. The amplitude is 
\begin{align}
\bar{u}_{l_q} ig_s \gamma^+ t^a \langle X| A^{-,a}(x)|N(s)\rangle \frac{-i}{\slashed{l}_q-\slashed{k}+i\epsilon}.
\end{align}
Then, take the modules square and perform the ensemble average of the nuclear medium, the differential scattering probability of the quark is (see the Appendix for details of derivation)
\begin{widetext}
\begin{align}
\frac{dP_q}{d^2\bfk} &= \frac{C_F}{d_A} \int_{r_0^+}^\infty dr^+  \rho_N(r^+, \bfb)  \frac{\alpha_s}{\bfk^2} \int
 \frac{dy^+ d\bfy^2}{2 \pi k^-} e^{ik^- y^+-i\bfk\cdot\bfy} \langle N|  F_i^-(y^+, \bfy) F^{-i}(0, 0)|N\rangle 
 \nonumber\\
&= \frac{C_F}{d_A}  \int_{r_0^+}^\infty dr^+  \rho_N(r^+, \bfb) \alpha_s \frac{\phi_g(x_g, \bfk^2)}{\bfk^2} \equiv  \int_{r_0^+}^\infty dr^+ \frac{d\Gamma_q}{d^2\bfk}
\label{eq:dNdkT2}
\end{align} 
\end{widetext}
where one has summed over the final state and averaged over the initial state spins and colors. Because of color confinement, the gluon field correlation only exists within a single nucleon. The expectation value over a nuclear wave function is reduced to the product of the expectation over a nucleon state $|N\rangle$ and the one-particle density $\rho_N(r^+, \bfb)$ along the path of the jet. The definition of TMD gluon distribution (neglecting the gauge link) is used in the last step, which defines $\phi_g(x_g, \bfk^2) =  4\pi x G_{\rm TMD}(x_g, \bfk^2)$. Hereafter, we will refer to $\phi_g$ as the TMD distribution in this study. The differential collision rate $d\Gamma/d^2\bfk$ is then directly related to the TMD gluon density distribution given by Eq. (\ref{eq:multiple_collisions}).
%For the collisional processes, $x_g = x_B \bfk^2/Q^2$ (see Appendix).
The calculation for a gluon can be obtained with the replacement $C_F\rightarrow C_A$ in Eq. (\ref{eq:dNdkT2}).

We consider the TMD gluon distribution $\phi_{g}(x_g, \bfk)$ approaches the Weizs\"acker-Williams distribution $1/\bfk^2$ at large $\bfk^2$, and introduce a saturation scale $Q_s$ that screens the infrared behavior, such that $\phi_{g}\sim 1/Q_s^2$ when $\bfk^2 \lesssim  Q_s^2$ \cite{Mueller:1999wm}.
Motivated by the picture of gluon saturation, in this version of eHIJING, we use a simple parametrization
\begin{align}
\alpha_s \phi_g(x_g, \bfk^2; Q_s^2) =  K \frac{(1-x_g)^n x_g^\lambda}{\bfk^2 + Q_s^2},
\label{eq:phig}
\end{align}
where $K$ is a constant factor, the powers $n$ and $\lambda$ parametrize the $x$ dependence. 
This is similar to the KLN model used for hadron production in proton-nucleus collision \cite{Kharzeev:2001gp,KHARZEEV2005609}. For the rest of the study, we will take $n=4$ and $\lambda=-0.25$ as given by \cite{KHARZEEV2005609}.
%so we neglect any explicit $Q^2$ dependence of the gluon TMD distribution. 
This simple model is sufficient for the study in this paper of in-medium jet fragmentation and momentum broadening. 
More sophisticated models can be implemented in the future. In particular, the scale evolution of $\phi_g(x,\bfk^2)$ is necessary for a more consistent study of jet modification over a large range of transverse momentum.

\begin{figure}
    \centering
    \includegraphics[width=\columnwidth]{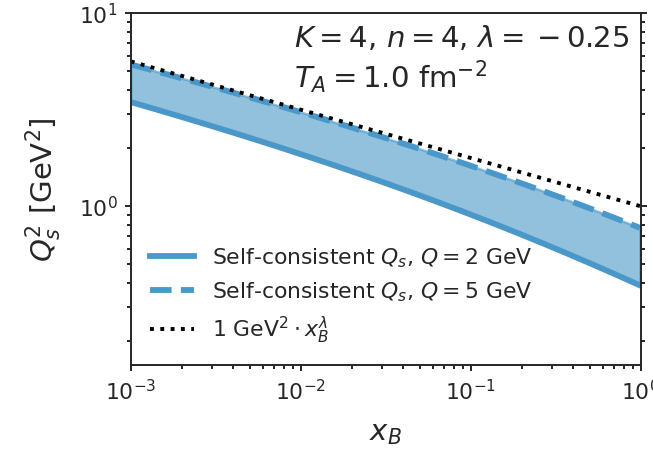}
    \caption{(Color online) The saturation scale $Q_s^2$ as a function of $x_B$ determined self-consistently from Eqs.~(\ref{eq:phig}) and (\ref{eq:self-consistent-Qs}) with parameters $K=4, n=4,\lambda=-0.25$ and
    the thickness function $T_A = 1.0$ fm$^{-2}$. The upper and lower edge of each band represents the variation of $2<Q<5$ GeV. The dashed line shows the asymptotic behavior of the saturation scale at small $x$. }
    \label{fig:Qs}
\end{figure}

With the above model for TMD gluon distribution, the saturation scale is determined by the self-consistent relation \cite{Zhang:2021tcc},
\begin{align}
Q_s^2 &= \frac{C_A}{d_A} T_A \int_0^{Q^2/x_B} d^2\bfk
\alpha_s \phi_g(x_g, \bfk^2; Q_s^2),
\label{eq:self-consistent-Qs}
\end{align}
where $T_A(r_0^+, \bfb) = \int_{r_0^+}^{\infty} dr^+ \rho(r^+, \bfb) $ is the thickness function of the nuclear matter passed by the jet. 
We allow the integration of $\bfk$ to go all the way up to the kinematic limit $Q^2/x_B$ when the gluon takes all the nucleon's momentum.  
In cases where the jet is produced close to the surface of the medium (i.e., $T_A$ is small), it is possible that the self-consistent equation results in $Q_s$ that is smaller than the QCD non-perturbative (NP) scale $\Lambda_{\rm QCD}$.
In this case, other NP effects will regulate the collinear behavior of Eq. (\ref{eq:phig}). Therefore, in eHIJING, a minimum value of $Q_{s,\rm min}=\Lambda_{\rm QCD} =0.25$ GeV is used.

We show the $x_B$, $Q^2$ dependence of the saturation scale in Fig. \ref{fig:Qs}. The evaluation uses $T_A=1.0\textrm{~fm}^{-2}$, comparable to the averaged nuclear thickness probed by DIS of a Pb nucleus.
The factor $K=4.0$ is chosen as it is found to give a reasonable description of the data in the result section.
Since $x_g=\frac{\bfk^2}{Q^2} x_B\ll 1$ at large $Q$, the $(1-x_g)^n$ term in Eq. (\ref{eq:phig}) is not very important, so $\phi_g \propto x_g^\lambda \propto x_B^\lambda$.
As a result, the self-consistent $Q_s^2$ is expected to scale as $x_B^{-\lambda}$. 
Such an asymptotic behavior is given by the black dotted line in Fig.~\ref{fig:Qs}.

\begin{figure}
    \centering
    \includegraphics[width=\columnwidth]{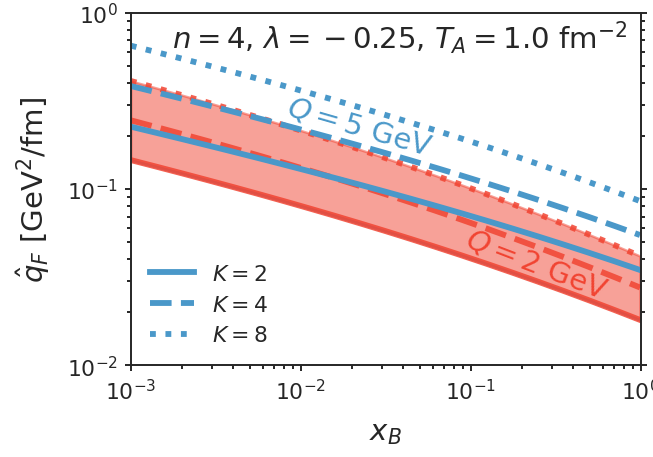}
    \caption{(Color online) The jet transport coefficient of a quark $\hat{q}_F =C_F Q_s^2/(C_A L)$ as a function of $x_B$ at two different values of $Q^2$. The solid, dashed, and dotted lines correspond to $K=2,4,8$.}
    \label{fig:qhat}
\end{figure}

Given the differential collision rates and the self-consistent saturation scale, one can compute the jet transport parameter $\hat{q}_R$. 
It is defined as the average momentum broadening per unit path length
\begin{align}
\hat{q}_R &\equiv \int_0^{Q^2/x_B} \bfk^2 \frac{d\Gamma_R}{d^2\bfk} d^2\bfk =  \frac{C_R}{C_A}\frac{Q_s^2}{L^+}.
\label{eq:qhat-2}
\end{align}
With the squared-transverse-momentum weighting, $\hat{q}_R$ is an infrared safe quantity so we extended the lower limit of $\bfk$ integration to zero.
The numerical value of $\hat{q}_R$ depends on the frame in which the path length is measured. So to avoid confusion, we will only quote its value in the rest frame of the nucleus, where $L^+$ is replaced by $L$.
The quark jet transport parameter as a function of $x_B$, $Q^2$, and the $K$ factor is shown in Fig. \ref{fig:qhat}.

\begin{figure}
    \centering
    \includegraphics[width=\columnwidth]{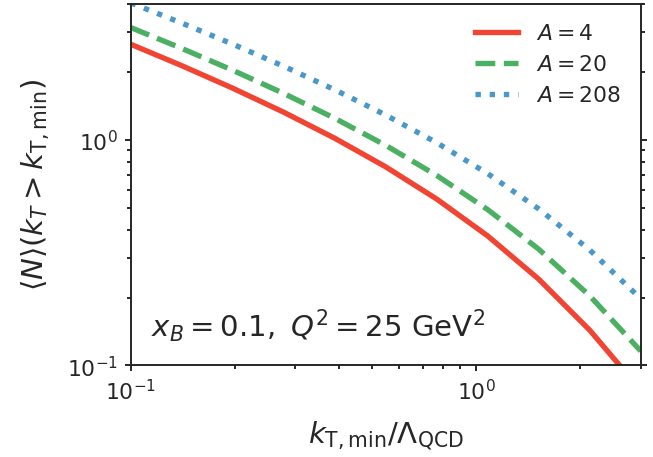}
    \caption{(Color online) The average number of multiple collisions with $k_T$ greater than $k_{T,\min}$. The calculation is for a quark produced with $x_B=0.1$ and $Q^2=25$ GeV$^2$ and uses the averaged $T_A$ of He (red), Ne (green), and Pb (blue line) nuclei.}
    \label{fig:Navg}
\end{figure}

Integrate the collision rate over the path length, one can define an average number of multiple collisions 
\begin{align}
\langle N\rangle (k_{T,\min}) = \int_{r_0^+}^\infty dr^+ \int_{|\bfk|_{\rm min}}^{|\bfk|_{\rm max}} d^2\bfk \frac{d\Gamma}{d^2\bfk}.
\end{align}
The kinematics gives the upper bound of the integration $|\bfk|_{\rm max} = \sqrt{Q^2/x_B}$. 
Unlike $\hat{q}_R$, the number of collisions is not an infrared safe quantity $\langle N\rangle$. Physically, we know that soft scatterings should be screened by non-perturbative effects, i.e., the nuclear matter is color-neutral at long distances.
So, a lower bound $|\bfk|_{\rm min}$ is introduced by hand to cut off the integration.
Fig. \ref{fig:Navg} shows the average number of scatterings as a function of $k_{T, \min}$ for light and heavy nuclei. As one can see, $\langle N\rangle$ increases when the infrared cut-off decreases: $\langle N\rangle_{\rm A=208}\approx 1$ when $k_{T, \min}=\Lambda_{\rm QCD}$ but increases to $4$ when $k_{T, \min}=0.1\Lambda_{\rm QCD}$.
However, physical observable effects, which are consequences of momentum broadening and parton energy loss, are not sensitive to $\langle N\rangle$ but $\hat{q}_R$, and the latter is an infrared safe quantity.
In the current version of eHIJING, the default choice is $k_{T, \min}=0.1\Lambda_{\rm QCD}$. One may consider increases $k_{T, \min}$ in the simulations as it avoids the sampling of ultra-soft scatterings. It does not affect the observable too much but can significantly improve the efficiency of the simulation.

If one assumes the nucleus is a dilute medium and subsequent scatterings are independent of one another. Then, the
event-by-event number of collisions $N$ follows a Poisson distribution 
\begin{align}
P_N = \frac{\langle N\rangle^N}{N!}e^{-\langle N\rangle},
\end{align}
given $\langle N\rangle$ is the averaged number of collisions.
Once $N$ is determined, the location of the scattering centers is randomly chosen along the path length. The transverse momentum exchange $\bfk$ of each collision is sampled according to Eq.~(\ref{eq:multiple_collisions}). 

\section{Medium-modified splitting functions}
\label{sec:model:rad}

\begin{figure}
    \centering
    \includegraphics[width=.8\columnwidth]{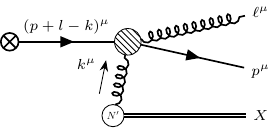}
    \caption{(Color online) The kinematic variables of a medium-induced real emission process.
    $k^\mu=(0, x_g p^-, \bfk)$ is the four-momentum of the gluon that collide with the jet parton. $q=p+\ell-k$ is the initial momentum of the hard parton.
    $\ell^\mu=[zq^+, \frac{\bfl^2}{2zq^+},\bfl]$ is the momentum of the radiated gluon. Finally, the recoiled quark's momentum is $p^\mu=[(1-z)q^+,\frac{(\bfk-\bfl)^2}{2(1-z)q^+},\bfk-\bfl]$. }
    \label{fig:cartoon2}
\end{figure}

There have been extensive studies on how jet-medium interactions modify parton splitting functions. They based on opacity expansion \cite{Gyulassy:2000er,Wiedemann:2000za,Ovanesyan:2011xy}, effective kinetic theory \cite{Arnold:2002ja,Arnold:2002zm}, BDMPS-Z formulation \cite{Baier:1998yf,Zakharov:1996fv,CaronHuot:2010bp} with harmonic oscillator approximation, and improved opacity expansion \cite{Mehtar-Tani:2019tvy,Mehtar-Tani:2019ygg,Barata:2020rdn}.
They differ on the assumptions about jet-medium collisions (``single-hard'' versus ``multiple-soft''), 
the kinematics of the radiative parton (full splitting versus soft radiation approximation), and properties of the medium  (``thin/dilute medium'' versus ``large/dense'' medium). 
An additional simplification on top of these is the twist expansion, where the resulting medium-modified splitting function is further expanded in powers of $1/Q^2$. In practice, the in-medium twist expansion is performed by investigating the calculation in powers of
$\bfk^2/\bfl^2$. $\bfk$ and $\bfl$ are the transverse momenta of multiple collisions and the radiated parton, respectively.
The kinematics variables for an in-medium parton splitting are illustrated in Fig. \ref{fig:cartoon2}.

\subsection{The generalized higher-twist (GHT) formula.}\label{sec:GHT}

For $e+A$ collisions,  we take the dilute limit and apply the results from a recent calculation of the medium-induced radiations from double parton scatterings in the generalized higher-twist approach \cite{Zhang:2021kym,Zhang:2021tcc}. In the Breit frame, a hard parton with finite transverse momentum $\bfp$ relative to the nucleus beam direction scatters with the virtual photon with momentum $q = [q^+, -Q^2/2q^+, {\bf 0}]$. The quark undergoes a second scattering with the nuclear target after the photon-quark hard scattering, exchanging a gluon with momentum $k$ and radiates a gluon with momentum $l = (zq^+, \bfl^2/2zq^+, \bfl)$.
The details of the expression for the radiative gluon spectra can be found in \cite{Zhang:2021kym,Zhang:2021tcc}, which can be summarized as 
\begin{align}
\frac{dN_{qg}^{\rm GHT}}{dz d^2\bfl} &= \frac{dN_{qg}^{\rm (0)}}{dz d^2\bfl} + \frac{dN_{qg}^{\rm (1)GHT}}{dz d^2\bfl}.
\end{align}
The first term is the vacuum-type contribution
\begin{align}
\frac{dN_{gq}^{\rm (0)}}{dz d^2\bfl} &= P^{(0)}_{gq} \frac{1}{\bfl^2}\\
P_{gq}^{(0)}(z) &= \frac{\alpha_s(\bfl^2) }{2\pi^2}C_F\frac{1+(1-z)^2}{z},
\end{align}
where $P^{(0)}_{ji}$ is the vacuum splitting function for parton $i$ to parton $j$ (plus another unspecified parton, either a quark or gluon).
The second term stands for the generalized twist-four contribution
\begin{align}
\frac{dN_{gq}^{\rm (1)GHT}}{dz d^2\bfl} &= P^{(0)}_{gq} \frac{1}{\bfl^2} \int_{r^+_0}^\infty dr^+\rho_N(r^+,\bfb) \nonumber \\
&\times \int_0^{Q^2/x_B} d^2\bfk \frac{C_A}{d_A}  \frac{ \alpha_s \phi_{g}(x_g, \bfk^2)}{\bfk^2}\nonumber\\
&\times \left[
\mathcal{N}_{g}^{q{\rm LPM}} + \mathcal{N}_{g}^{g{\rm LPM}} + \mathcal{N}_g^{\rm nonLPM}\right].
\label{eq:1to2_full}
\end{align}
%where we choose the convention that the splitting function is defined with the factor $\alpha_s/(2\pi)$.
The precise definition of the interference terms $\mathcal{N}_{g}^{g{\rm LPM}}$, $\mathcal{N}_{g}^{q{\rm LPM}}$ and $\mathcal{N}_g^{\rm nonLPM}$ can be found in Ref. \cite{Zhang:2021tcc}.
$\mathcal{N}_{g}^{q{\rm LPM}}$ is contributed by rescatterings between the quark and the medium gluon, and $\mathcal{N}_{g}^{g{\rm LPM}}$ from rescatterings of the radiated gluon with the medium gluon. Both of these terms are enhanced by the nuclear size and contain the so-called Landau-Pomeranchuk-Migdal (LPM) interference factor, which is to be explained shortly. 
The term $\mathcal{N}_g^{\rm nonLPM}$ is not enhanced by the nuclear size, and its contribution is suppressed by the ultraviolet cut-off of the radiation.
The interference factors are integrated over the collisional cross-section ($\propto \alpha_s \phi_g /\bfk^2$), and the nucleon density distribution $\rho_N$ along the path length. Therefore, the obtained semi-analytic form shown in Eq. (\ref{eq:1to2_full}) should be understood as an average one over the multiple collisions inside the nuclei. However, in eHIJING, the chain of multiple collisions is sampled randomly. Later in this section, we will propose a method to implement a stochastic version of Eq. (\ref{eq:1to2_full}) that is consistent with the sampled collisions.

\subsection{The soft gluon approximation (SGA) of the GHT formula}
\label{sec:GHT_SGA}
To simplify the simulations, we take the soft gluon emission approximation of Eq. (\ref{eq:1to2_full}).
In this limit, both $\mathcal{N}_{g}^{q{\rm LPM}}$ and   
$\mathcal{N}_g^{\rm nonLPM}$ are negligible~\cite{Zhang:2021kym,Zhang:2021tcc}. In the remaining term $\mathcal{N}_{g}^{g{\rm LPM}}$, the soft gluon limit further decouples the formula from the transverse momentum of the initial state quark ($\bfq$). 
Therefore, the transverse momentum recoil from medium-induced gluon radiations can be sampled independently of the initial state quark's transverse momentum \footnote{The relaxation of the approximation $z\ll 1$ may introduce a non-trivial correlation between quark TMD PDF and the modified fragmentation function.}.

In the soft-gluon limit, the splitting function (summation of the vacuum and medium-induced contribution) simplifies to,
\begin{align}
&\frac{dN_{gq}^{\rm GHT}}{dz d\bfl^2} 
\approx  P_{gq}^{(0)}(z)\frac{1}{\bfl^2} +  
P_{gq}^{(0)}(z)\frac{1}{\bfl^2}\int_{r^+_0}^{\infty} dr^+ \rho_N(r^+, \bfb)  \nonumber\\
& \int_0^{Q^2/x_B} \frac{C_A}{d_A}\frac{\alpha_s \phi_{g}}{\bfk^2} \frac{2\bfk\cdot \bfl}{(\bfl-\bfk)^2}
\Phi\left(\frac{\Delta r^+}{\tau_f}\right)d^2\bfk ,
\label{eq:1to2_soft_and_static_q2qg}
\end{align}
where $\Delta r^+ = r^+-r^+_0$ is the location of the scattering center relative to the hard production vertex. $\tau_f$ is the formation time of the medium-induced splitting
\begin{align}
\tau_f  =\frac{2z(1-z)q^+}{(\bfl-\bfk)^2}.
\end{align}
The appearance of the interference phase factor 
\begin{align}
\Phi(\Delta r^+/\tau_f) = 1-\cos(\Delta r^+/\tau_f)
\end{align}
in the splitting function qualitatively changes the behavior 
of the medium-induced part of the splitting function compared to the vacuum part.
For emissions with long formation times $\tau_f\gg \Delta r^+ \sim L^+$, the phase factor strongly suppresses the medium-induced contributions, which is known as the QCD analog of the Landau-Pomeranchuk-Migdal (LPM) interference effect. In the following discussion, we will refer to the limit $\tau_f\gg L^+$ as the coherent limit.
In the other limit (incoherent) where $\tau_f \ll L$, the cosine factor averages to zero under the $\Delta r^+$ integration, and effectively $\Phi(\Delta^+/\tau_f\gg 1) = 1$. 
Then, one finds that in the incoherent limit, the $z$ dependence of the medium-induced part is the same as the vacuum-splitting function.

Similarly, the SGA of the medium-modified $g\rightarrow g$ splitting function is 
\begin{align}
&\frac{dN_{gg}^{\rm GHT}}{dz d\bfl^2} 
\approx  P_{gg}^{(0)}(z)\frac{1}{\bfl^2} +  
P_{gg}^{(0)}(z)\frac{1}{\bfl^2}\int_{r^+_0}^{\infty} dr^+ \rho_N(r^+, \bfb)  \nonumber\\
& \int_0^{Q^2/x_B} \frac{C_A}{d_A}\frac{\alpha_s \phi_{g}}{\bfk^2} \frac{2\bfk\cdot \bfl}{(\bfl-\bfk)^2}
\Phi\left(\frac{\Delta r^+}{\tau_f}\right)d^2\bfk ,
\label{eq:1to2_soft_and_static_g2gg}
\end{align}
with the vacuum splitting function
\begin{align}
P_{gg}^{(0)}&= \frac{\alpha_s(\bfl^2) }{2\pi}C_A\left[ \frac{1+(1-z)^3}{z} +  \frac{1+z^3}{1-z}\right].
\end{align}
Here $P_{gg}^{(0)}$ is decomposed into two pieces to be compatible with Pythia8's color dipole picture in the $g\rightarrow g+g$ process \cite{Gustafson:1987rq}.

To make a connection with past works, e.g., the GLV formula widely used in jet quenching phenomenology \cite{Gyulassy:2000er}, one replaces the TMD gluon distribution with a  screened Coulomb potential (also known as the Gyulassy-Wang model \cite{Wang:1994fx}) to model the interaction between the jet parton and the nucleus  
\begin{align}
\frac{\alpha_s \phi_g(x_g, \bfk^2)}{\bfk^2} \rightarrow \frac{4\pi\alpha_s^2 \sum_T C_T f_T}{(\bfk^2+m^2)^2},
\label{eq:GWmodel}
\end{align}
where $C_T$ and $f_T$ are the color factor and the probability of finding a color source of representation $T$ within the nucleon.
Then, Eqs. (\ref{eq:1to2_soft_and_static_q2qg}) and (\ref{eq:1to2_soft_and_static_g2gg}) will reduce to the GLV formula used in \cite{Gyulassy:2000er}. In the rest of this work, we will continue to use the TMD gluon distribution to parameterize the jet-medium interaction.

\subsection{The reduction to the higher-twist (HT) formula under SGA} 
\label{sec:HT_SGA}
Another well-applied method to compute medium modifications is the higher-twist approach \cite{Wang:2001ifa}, where one expands the calculation in powers of $1/\bfl^2$ and keeps the twist-four contributions.
With such expansions, Eqs.~(\ref{eq:1to2_soft_and_static_q2qg}) and (\ref{eq:1to2_soft_and_static_g2gg}) become
\begin{align}
& 
\begingroup
\renewcommand*{\arraystretch}{2}
\begin{Bmatrix}
\frac{dN_{gq}^{\textrm{HT}}}{dz d\bfl^2} \\
\frac{dN_{gg}^{\textrm{HT}}}{dz d\bfl^2}
\end{Bmatrix} 
\endgroup
\begingroup
\renewcommand*{\arraystretch}{1.5}
= \begin{Bmatrix}
P_{gq}^0(z)\\
P_{gg}^0(z)
\end{Bmatrix} 
\endgroup
 \frac{1}{\bfl^2} + \nonumber\\
&  
\begingroup
\renewcommand*{\arraystretch}{1.5}
\begin{Bmatrix}
P_{qg}^0(z)\\
P_{gg}^0(z)
\end{Bmatrix} 
\endgroup
\int_{r^+_0}^\infty dr^+  \frac{2\hat{q}_A^{\rm rad}}{\bfl^4} \Phi\left(\frac{\Delta r^+}{\tau_f'}\right) + \mathcal{O}\left(\frac{1}{\bfl^6}\right),
\label{eq:1to2_soft_and_static_and_collinear}
\end{align}
with the twist-expanded formation time $\tau_f'$ being~\footnote{Note that one should not expand $\tau_f$ in this manner when either $z$ or $1-z$ goes to zero, i.e., when $\Phi$ is highly oscillating. Here, we are following the practice of the original derivation of the Higher Twist results. }
\begin{align}
\tau_f'  =\frac{2z(1-z)q^+}{\bfl^2}.
\end{align}
The radiative transport parameter is defined as,
\begin{align}
\hat{q}_A^{\rm rad} = \rho_N
 \int_0^{\bfl^2} \frac{C_A}{d_A} \alpha_s \phi_{g}(x_g, \bfk^2) \pi d\bfk^2 ,
 \label{eq:qhat_rad}
\end{align}
which should not be confused with the collisional transport parameter $\hat{q}_R$, since the range of $\bfk$ integration is, in general, different from the one used in Eqs. (\ref{eq:self-consistent-Qs}) and (\label{eq:qhat-2}) for calculating $\hat{q}_R$.
In the original derivation, one assumes that  $\bfk^2/\bfl^2$ is a small number and performs the expansion before the $\bfk$ integration, then, a consistent integration range should be chosen as $0 < |\bfk| < |\bfl|$.

However, we would like to point out that there is a certain level of ambiguity in the definition of the radiative transport parameter in Eq. (\ref{eq:qhat_rad}).
For example, one can consider another way of obtaining the higher-twist expansion from the GHT formula, where one performs the $1/\bfl^2$ expansion after the $\bfk$ integration.
To do so, one makes a change of variable $\bfl \rightarrow \bfl+\bfk$ in Eqs. (\ref{eq:1to2_soft_and_static_q2qg}) and (\ref{eq:1to2_soft_and_static_g2gg}) and neglecting boundary terms power suppressed by $1/Q^2$ that are not enhanced by the nuclear size.
The $\bfk$ integration can then be worked out as 
\begin{align}
\frac{dN_{ji}^{\rm(1) GHT}}{dz d\bfl^2} &= P_{ji}(z) \frac{1}{\bfl^2} \int_{r_0^+}^\infty dr^+  \rho_N \int_0^{Q^2/x_B} d^2\bfk   \nonumber\\
&\frac{C_A}{d_A}\frac{\alpha_s \phi_{g}}{\bfk^2}\frac{2\bfk\cdot \bfl}{(\bfl-\bfk)^2} \Phi\left(\frac{(\bfl-\bfk)^2\Delta r^+}{2z(1-z)q^+}\right) \nonumber\\
&= P_{ji}(z) \int_{r_0^+}^\infty dr^+ \frac{2}{\bfl^4}  \bfl^2\int_{\bfl^2}^{Q^2/x_B}     \pi d\bfk^2 \nonumber\\
& \rho_N\frac{C_A}{d_A}\frac{\alpha_s \phi_{g}}{\bfk^2} \Phi\left(\frac{\Delta r^+}{\tau_f'}\right) +\mathcal{O}\left(\frac{1}{Q^2}\right).
\end{align}
In comparison with the form of the higher-twist formula in Eq. (\ref{eq:1to2_soft_and_static_and_collinear}), it turns out that the effective radiative transport parameter should be defined as
\begin{equation}
\hat{q}_A^{\rm rad'} = \bfl^2 \rho_N \int_{\bfl^2}^{Q^2/x_B}    \frac{C_A}{d_A}\frac{\alpha_s \phi_{g}(x_g, \bfk^2)}{\bfk^2} \pi d\bfk^2,
\end{equation}
and then expand to the leading power of $1/\bfl^2$.
For example, if one uses fixed coupling and applies the Gyulassy-Wang (GW) screened potential model expressed in Eq. (\ref{eq:GWmodel}), it can be shown \cite{Cao:2020wlm} that 
\begin{align}
\hat{q}_{\rm GW}^{\rm rad} &= 
\sum_T \alpha_s^2 \rho_N \frac{C_A C_T}{d_A} 4\pi^2 f_T \ln\frac{\bfl^2}{m^2} + \mathcal{O}\left(\frac{m^2}{\bfl^2}\right), \\
\hat{q}_{\rm GW}^{\rm rad '} &=  \sum_T \alpha_s^2 \rho_N \frac{C_A C_T}{d_A} 4\pi^2 f_T + \mathcal{O}\left(\frac{m^2}{\bfl^2}, \frac{\bfl^2}{Q^2}\right).
\end{align}
So, there is an $\ln\frac{\bfl^2}{m^2}$ ambiguity when relating the radiative transport parameter to the more fundamental input of jet-medium interaction. 
In this paper, we will use the first choice in Eq. (\ref{eq:qhat_rad}).

The medium-induced splitting function takes a complicated form, but its primary physical effect is intuitively --- the medium-induced radiative energy loss. 
One can compute the $z$-weighted integration of $dN_{ji}^{(1)}$ and arrive at the averaged loss in energy. For example, using the GW model, this is 
\begin{equation}
\Delta z \propto \alpha_s \frac{\hat{q}_{GW}^{\rm rad'} L L^+}{q^+} \ln\frac{q^+/L^+}{m^2}
\end{equation}
in the GHT approach. The radiative energy loss in the medium is proportional to the quadratic power of the path length is a well-known behavior of QCD. This is similar for the HT approach but the log enhancement factor is different.

\subsection{A numerical comparison of higher-twist and the generalized higher-twist formula.}

In eHIJING, we will implement both ways of treating the modified parton splitting using either 1) the generalized higher-twist (GHT) formula in Eqs.~(\ref{eq:1to2_soft_and_static_q2qg}) and (\ref{eq:1to2_soft_and_static_g2gg}), or 2) the higher-twist (HT) formula in Eq. (\ref{eq:1to2_soft_and_static_and_collinear}).
Even though here the HT formula is obtained from the twist-expansion of the GHT formula, we still consider it to be valuable to implement both choices in eHIJING for two reasons
\begin{enumerate}
\item The first purpose is to provide a formal benchmark of the two approaches since both have been widely used in the past in heavy-ion collisions and $e$+$A$ phenomenology. 
\item Second, the GHT formula relies on the assumption that the microscopic interactions between the jet parton and nuclear medium can be treated as a perturbative forward scattering cross-section.
In the HT approach, the effects of multiple scatterings are absorbed into the $\hat{q}_A^{\rm rad}$ parameter.
For phenomenology, one can treat $\hat{q}^{\rm rad}$ as an effective non-perturbative parameter, which can, in principle, avoid the details of the microscopic modeling of the jet medium interactions.
\end{enumerate}

\begin{figure}
    \centering
    \includegraphics[width=\columnwidth]{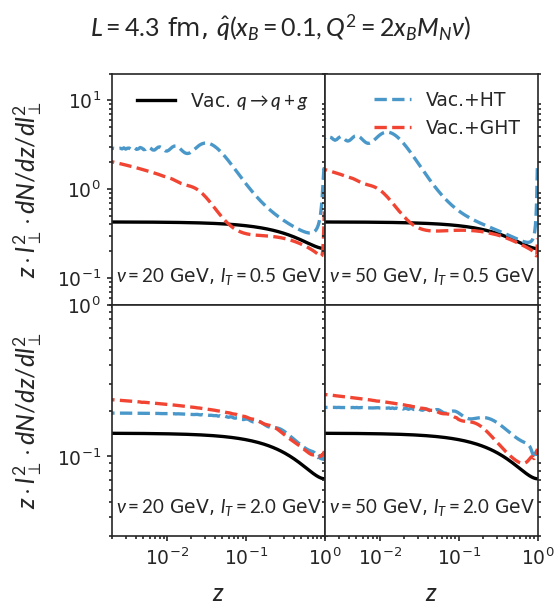}
    \caption{(Color online) In-medium splitting functions in the generalized higher-twist approach (red) and the higher-twist approach (blue) for two different values of transverse momentum $|\bfl |$ (rows) and different quark energy $\nu$ in the nuclear rest frame (columns). Black lines are vacuum-splitting functions as a reference.}
    \label{fig:compare_HT_Gen_2}
\end{figure}

To visualize the differences between the GHT and HT formula when using the same microscopic input $\alpha_s \phi_g(x_g,\bfk^2)$, in Fig. \ref{fig:compare_HT_Gen_2}, we compare the spectra of gluon emission $z \bfl^2 dN_{gq}/dz/d\bfl^2$ from a quark.
The path length is chosen as $L=4.3$ fm -- the averaged path length for a heavy nucleus with $A\approx 100$.
In each panel, the black lines are the vacuum emission spectrum $dN_{gq}/dz/d\bfl^2=P_{gq}^{(0)}(z)/\bfl^2$ as a function of $z$.
The red (blue) dashed lines represent the sum of the vacuum and medium-induced splitting functions using the GHT (HT) formula.
The top and bottom rows are comparisons of a small and large transverse momentum of the gluon ($|\bfl|=0.5$ GeV and $|\bfl|=2.0$ GeV).
The left and right columns vary the quark energy in the rest frame of the nucleus from $\nu=20$ GeV to $\nu=50$ GeV.

Because the HT formula is obtained based on $1/\bfl^2$ expansion, it is not surprising that the difference between GHT and HT is relatively small at large $|\bfl|$ (bottom row).
Furthermore, at larger $\bfl$ and smaller $\nu$ (the bottom left panel), the ratio $L/\tau_f \sim L/\tau_f' \sim \frac{\bfl^2 L}{2z(1-z)\nu} \gg 1$. This means that the LPM phase factor $\Phi\approx 1$ and the $z$-dependence of the medium-induced contribution are the same as the vacuum ones, which is confirmed by noticing that the three lines in the bottom left panel have a similar shape.

At small $|\bfl|$ and large $\nu$ (the top right panel), the ratio $L/\tau_f\ll 1$ unless $z\ll 1$ or $(1-z)\ll 1$. Therefore, the medium-induced collinear emissions are completely suppressed, and the ``vac.+HT'' and ``vac.+GHT'' curves almost overlap with the vacuum curve near $z=1/2$. 

Finally, at small $|\bfl|$, the higher-twist formula leads to much stronger modifications especially at large $z$, while the correction in the generalized higher-twist is smaller.
As a result, with the same microscopic input $\alpha_s \phi_g(x_g, \bfk^2)$, the HT approach induces more collinear radiations than the GHT approach and also leads to a stronger parton energy loss than the latter.
We will discuss the phenomenological impact of such differences in the result section.

\subsection{A stochastic version of the medium-modified splitting functions}

It is mentioned at the end of Sec. \ref{sec:GHT} that the modified splitting functions presented in Secs. \ref{sec:GHT_SGA} and \ref{sec:HT_SGA} are inclusive over the number of the multiple collisions and their kinematics. This is evident by noticing that Eqs. (\ref{eq:1to2_soft_and_static_q2qg}), (\ref{eq:1to2_soft_and_static_g2gg}), and (\ref{eq:1to2_soft_and_static_and_collinear}) are integrated over the location $r^+$ and the transverse momentum $\bfk$ of the multiple collisions. On the other hand, the eHIJING event generator samples multiple collisions stochastically according to the procedure described at the end of Sec. \ref{sec:model:coll}. 

\begin{figure}
\centering
\includegraphics[width=.8\columnwidth]{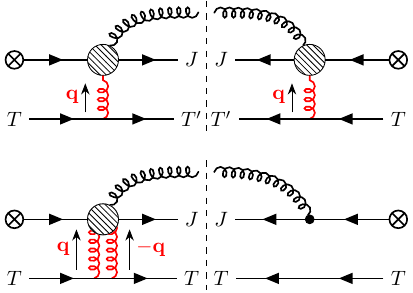}
\caption{(Color online) Schematic plot of two types of contribution to the medium-modified splitting function. Top: diagrams that involve a real collision with a recoiled target $T\rightarrow T'$. Bottom: diagrams that contain virtual interaction with the target. Note that in this case, the state of the target is unchanged $T\rightarrow T$.}
\label{fig:unitarity}
\end{figure}
Ideally, we would like to modify the splitting function consistently with the stochastic sample of the multiple collisions.
This guarantees that if there are not any parton-medium interactions sampled in a particular event, the splitting function reduces to the vacuum one.
We have to admit that, at the moment, we lack the full knowledge of the proper way to sample medium-induced radiative partons, multiple collisions, and the recoiled medium partons consistently.
The fundamental problem is that 
the medium-induced part of the splitting function in Eqs.~(\ref{eq:1to2_soft_and_static_q2qg}) and (\ref{eq:1to2_soft_and_static_g2gg}) contains two classes of contributions: 1) the collision between jet and medium (as illustrated on the left of Fig. \ref{fig:unitarity}), and 2) the unitarity correction --- interference between vacuum emission diagram and the diagram with double-gluon exchange between jet and the medium without net momentum transfer (the right of Fig. \ref{fig:unitarity}).
If one intends to sample both the radiative gluon and the exchanged gluon $\bfk$ calculation, only the first 
class of contributions should be associated with a real collision with the nucleus.  

It is not the purpose of this work to solve this problem right away, but we would like to propose an ansatz to construct a stochastic version of the medium-modified splitting functions that contains reasonable correlation with the randomly sampled multiple collisions.

For a given hard parton with light-cone momentum $yq^+$ in an event, suppose the number of multiple collisions $N$ and the location and transverse momentum have been sampled, which gives the following sequence of collision information
\begin{align}
\{(r_1^+, {\bfk}_1), \cdots, (r_i^+, {\bfk}_i), \cdots, (r^+_N,{\bfk}_N)\}.
\label{eq:sequence}
\end{align}
They form an important sampling of the following measure
\begin{equation}
\int_{r^+_0}^{\infty} dr^+ \rho_N(r^+,\bfb) \int_0^{\frac{Q^2}{x_B}} d^2\bfk \frac{C_R}{d_A}\frac{\alpha_s \phi_{g}(x_g, \bfk)}{\bfk^2},
\end{equation}
which also appears in the computation of the medium-modified splitting functions.
Therefore, we propose a stochastic version of the GHT splitting functions by replacing the corresponding measure in Eqs. (\ref{eq:1to2_soft_and_static_q2qg}) and (\ref{eq:1to2_soft_and_static_g2gg}) with the sum of the random samples
\begin{align}
&\begingroup
\renewcommand*{\arraystretch}{2}
\begin{Bmatrix}
\frac{dN_{gq}^{\textrm{GHT}}}{dz d\bfl^2} \\
\frac{dN_{gg}^{\textrm{GHT}}}{dz d\bfl^2}
\end{Bmatrix} 
\endgroup
\Rightarrow
\begingroup
\renewcommand*{\arraystretch}{1.5}
\begin{Bmatrix}
P_{gq}^0(z)\\
P_{gg}^0(z)
\end{Bmatrix} 
\endgroup \frac{1}{\bfl^2} + 
\begingroup
\renewcommand*{\arraystretch}{1.5}
\begin{Bmatrix}
P_{gq}^0(z) \frac{C_A}{C_F}\\
P_{gg}^0(z)
\end{Bmatrix} \frac{1}{\bfl^2}\times
\endgroup
 \nonumber\\
& 
 \sum_{i=1}^N \frac{2{\bfk}_i\cdot \bfl}{(\bfl-{\bfk}_i)^2}\left. \Phi\left(\frac{(\bfl-{\bfk}_i)^2 (r_i^+-r_0^+)}{2z(1-z)yq^+} \right) \right\}.
 \label{eq:GHT-stochastic}
\end{align}
Similarly, the stochastic version of the HT splitting functions is
\begin{align}
& 
\begingroup
\renewcommand*{\arraystretch}{2}
\begin{Bmatrix}
\frac{dN_{gq}^{\textrm{HT}}}{dz d\bfl^2} \\
\frac{dN_{gg}^{\textrm{HT}}}{dz d\bfl^2}
\end{Bmatrix} 
\endgroup
\begingroup
\renewcommand*{\arraystretch}{1.5}
= \begin{Bmatrix}
P_{gq}^0(z)\\
P_{gg}^0(z)
\end{Bmatrix} 
\endgroup
 \frac{1}{\bfl^2} + \begingroup
\renewcommand*{\arraystretch}{1.5}
\begin{Bmatrix}
P_{qg}^0(z)\frac{C_A}{C_F}\\
P_{gg}^0(z)
\end{Bmatrix} 
\endgroup \frac{1}{\bfl^2}\times 
\nonumber\\
&  \sum_{i=1}^N\Theta(\bfl^2-{\bfk^2}_i)\frac{2{\bfk^2}_i}{\bfl^2} \Phi\left(\frac{\bfl^2(r^+_i-r^+_0)}{2z(1-z)yq^+}\right).
\label{eq:HT-stochastic}
\end{align}
These modified splitting functions are correlated with multiple collisions. In particular, for events with $N=0$ or with a tiny collisional transverse momentum transfer, Eqs. (\ref{eq:GHT-stochastic}) and (\ref{eq:HT-stochastic}) reduces to $P^{(0)}_{ji}(z)$. 
It is also true that a parton with smaller medium modifications to its splitting functions, and thus less parton energy loss, is also subjected to a lesser amount of transverse momentum broadening. 
This correlation is the key to understanding the decrease of hadron transverse momentum broadening near the production threshold in the result section.

\section{Implementing modified splitting functions in parton evolution and fragmentation}
\label{sec:model:evo}
In the vacuum, radiative corrections are enhanced by the logarithm of the phase-space of the emission $Q_0^2<\bfl^2<Q^2$, with $Q_0\gtrsim \Lambda_{\rm QCD}$ being a separation scale between the perturbative and non-perturbative physics.
In the leading-log approximation, terms like $[\alpha_s(Q_0^2) \ln(Q^2/Q_0^2)]^n$ are resumed and leads to the Dokshitzer-Gribov-Lipatov-Altarelli-Parisi (DGLAP) \cite{Gribov:1972ri,Dokshitzer:1977sg,Altarelli:1977zs} evolution of the fragmentation functions with the energy scale. 
In a parton shower picture, it is viewed as the scale evolution of the parton spectrum from virtuality comparable to $Q^2$ down to  $Q_0^2$. Thus, subsequent splittings are ordered in decreasing virtuality.

In Pythia8, the ordering quantity is chosen as the transverse momentum $|\bfl|$ generated in each splitting. For a parton $i$ produced with ${\bfl}_1$ in the previous splitting, to generate the next splitting in the channel $i\rightarrow j$, the algorithm iteratively samples the Sudakov form factor $\Delta_{ji}^{(0)}({\bfl}_2, {\bfl}_1)$, i.e., the probability of no emissions between the kinematic regions ${\bfl}_2<\bfl<{\bfl}_1$,
\begin{align}
r = \Delta_{ji}^{(0)}({\bfl}_2, {\bfl}_1) = e^{-\langle N_{ji}^{(0)}\rangle\left({\bfl}_2, {\bfl}_1\right) }, 
\label{eq:Sudakov_vac}
\end{align}
where $r$ is a sample of a random number uniformly distributed between 0 and 1, and $\langle N_{ji}^{(0)}\rangle\left({\bfl}_2, {\bfl}_1\right) $ the average number of vacuum radiations within the given kinematic regions,
\begin{align}
 \langle N_{ji}^{(0)}\rangle({\bfl^2}_2,{\bfl^2}_1) =& \int_{z_{\rm min}(\ell_m)}^{z_{\rm max}(\ell_m)}  P^{(0)}_{ji} (z) dz \nonumber\\
 &\int_{{\bfl^2}_2}^{{\bfl^2}_1}\frac{\alpha_s(\bfl^2)}{2\pi^2} \frac{d^2\bfl}{\bfl^2}.
 \label{eq:Nrad_vac}
\end{align}
Here the lower bound $z_{\rm min}$ and upper bound $z_{\rm max}$ of the momentum fraction are determined by the minimum possible transverse momentum ($\ell_m=0.5$ GeV in the default Pythia8 setting).
The solution of Eqs. (\ref{eq:Sudakov_vac}) and (\ref{eq:Nrad_vac}) determines the transverse momentum ${\bfl}_2$ for the next splitting $i\rightarrow j$.
Then, the momentum fraction $z$ is sampled according to the splitting function $P_{ji}^{(0)}(z)$ between $z_{\rm min}(\ell_m)<z<z_{\rm max}(\ell_m)$. Finally, samples falling outside of the physical domain  $z_{\rm min}(|\bfl|_2)<z<z_{\rm max}(|\bfl|_2)$ are rejected. This way, the full kinematics of the splitting is given by $z$ and ${\bfl}_2$.

In Pythia8, the upper bound of the first emission is $|\bfl|\sim Q$ and the whole interactive procedure terminates when the sampled $|\bfl|$ is below $l_m$.
At this point, the perturbative parton shower gives way to the non-perturbative modeling of hadronization. 
Pythia8 uses the Lund string hadronization model \cite{Sjostrand:2014zea}. The color-neutral system of partons forms strings according to the flow of color. Then, the string-breaking mechanism iteratively samples hadron from the string system, with hadronic decays applied afterward. In $e$+$p$ reactions, it is important to include the proton remnants such that the combined system of parton shower and remnants is color neutral.

In the $e$+$A$ reactions, due to the emergence of new energy scales, such as $Q_s^2$ and $\nu/L$ \cite{Ke:2023ixa}, from dynamical effects when jet parton propagates in a finite-size medium, the parton shower dynamics is divided into different stages. 
For medium-induced contribution that involves a large virtuality $\bfl^2 \gg Q_s^2$, the multiple emissions are generated in a virtuality/scale ordered shower, but with the set of medium modified splitting functions  $\frac{dN}{dz d\bfl^2}=\frac{dN^{(0)}}{dz d\bfl^2}+\frac{dN^{(1)}}{dz d\bfl^2}$ \cite{Chang:2014fba}, which are developed in Eq. (\ref{eq:GHT-stochastic}) or Eq. (\ref{eq:HT-stochastic}).
An emission is sampled using the modified Sudakov factor
\begin{align}
r = \Delta_{ji}({\bfl}_2, {\bfl}_1) &= e^{- \langle N_{ji}\left({\bfl}_2, {\bfl}_1\right) \rangle}, 
\label{eq:Sudakov_med}\\
\langle N_{ji}({\bfl}_2, {\bfl}_1)\rangle &= \int_{z_{\rm min}(\ell_m)}^{z_{\rm max}(\ell_m)} 
 dz \int_{{\bfl^2}_2}^{{\bfl^2}_1} d\bfl^2 \nonumber\\
 &\left[\frac{dN^{(0)}}{dz d\bfl^2} + \frac{dN^{(1)}}{dz d\bfl^2}\Theta(Q_s^2<\bfl^2)\right]
 \label{eq:Nrad_med}
\end{align}
The modified Sudakov is added by eHIJING to the Pythia8 $k_T$-ordered shower. The virtuality shower still runs from $\bfl\sim Q$ down to $\bfl = \ell_m$ but the medium contribution is switched off from the Sudakov factor if $\bfl^2<Q_s^2$.

If $\bfl^2$ becomes comparable or smaller than the screening/saturation scale $Q_s^2$, the medium contribution is no longer enhanced by the logarithm of phase space but by the length of propagation in the matter.
Therefore, we implemented a time-ordered shower for parton emissions for the medium-induced contribution with $\bfl^2<Q_s^2$. For this purpose, we solve the Sudakov form factor of no-emission between the formation time $\tau_2<\tau_f<\tau_1$
\begin{align}
r = \Delta_{ji}(\tau_2, \tau_1) = e^{-\langle N_{ji}^{(1)} \rangle(\tau_2, \tau_1)},
\label{eq:sudakov-tau}
\end{align}
where $r$ is a random number uniformly distributed from 0 to 1, and $\Delta_{ji}(\tau_2, \tau_1)$ is the Sudakov factor, i.e., the probability of vetoing radiation within formation time between $\tau_2$ and $\tau_1$. The total number of gluons $N_g$ between the two times is  
\begin{align}
  \langle N_{ji}^{(1))}\rangle(\tau_2, \tau_1) =& \int_{\Lambda_{\rm QCD}^2}^{Q_s^2} \frac{d\bfl^2}{\bfl^2} \int_0^1 dz \frac{dN_{ji}^{(1)}}{dz d\bfl^2}\nonumber\\
  &\Theta\left(\tau_1<\frac{2z(1-z)p^+}{\bfl^2}<\tau_2\right).
\end{align}
The starting time of the evolution is chosen as $\tau_1=1/Q$. The solution to Eq. (\ref{eq:sudakov-tau}) then determines the formation time of the next emission. Then the sampling of $z$ and the azimuthal angle $\phi$ determine the kinematics of the emission. The maximum time is cut off at $t_{\rm max} = E/\Lambda_{\rm QCD}^2$ --- the typical timescale of hadronization.

\section{Remnant, collisional energy loss, and hadronization}\label{sec:model:remnant}
The Lund string hadronization model only applies to a color-singlet multi-parton system. In $e$+$p$ or $pp$ collisions, the entire system of the hard process plus the proton remnant remains color-neutral. For example, when a gluon from the proton participates in the hard process, the proton remnant is modeled by a quark plus a diquark that carries the corrected color and flavor information.

In $e$+$A$ collisions, in addition to the nucleon remnant from the primary hard process, there are also remnants from the multiple parton-medium scatterings.
When calculating the jet sector, the ``+'' component of the gluon and the nucleon are neglected in Eqs. (\ref{eq:jet-side-k}) and (\ref{eq:jet-side-p}). However, it is critical to keep $k^+$ and $p^+$ finite when discussing the nucleon remnant. Since the invariant mass square $m_R^2$ of the nucleon remnant mass is positive,
\begin{align}
m_R^2 &= (p-k)^2 = (1-x_g)\left(m_p^2 - k^+p^-\right) - \bfk^2,\\
k^+ &= \frac{(1-x_g)m_p^2-m_R^2-\bfk^2}{(1-x_g)p^-},
\label{eq:remnant-momentum}
\end{align}
where in the last line, we have used $x_g\ll1$ and $\bfk^2\sim Q_s^2$ for a typical collision.
If we assume that the remnant carries the baryon number, then $m_R$ should at least be comparable or greater than the mass of a proton $m_R\gtrsim m_p$. 
Using $x_g\ll 1$ and $\bfk^2\sim Q_s^2$ for typical collisions, one finds that $k^+\sim Q_s^2/p^- < 0$. Therefore, to guarantee the energy-momentum conservation on the target-going side, the jet parton must lose a fraction of its energy $\frac{k^+}{q^+}\Gamma L \sim Q_s^2/(2(1-x_B)M\nu)$ via collisional process --- referred as collisional energy loss. In the nuclear rest frame, the collisional energy loss is of order $\frac{Q_s^2}{M}$, which is subleading to the radiative energy loss $\Delta E\sim \alpha_s Q_s^2 L \ln\frac{\nu/L}{Q_s^2}  \sim \frac{\alpha_s Q_s^2}{M} A^{1/3}\ln\frac{\nu/L}{Q_s^2}$ when either the energy of the parton or the medium size is large. However, it may be important for fixed-target experiments. For this reason, we provide an option to turn on collisional energy loss using the simplified formula 
\begin{align}
k^+ = -\frac{\bfk^2}{p_q^-}
\end{align}
Its phenomenological effect is also demonstrated in the result section.

To construct the remnant, the four-momentum of the quark $k_q$ and diquark $k_{qq}$ before the multiple collision is sampled isotropically in the rest frame of the proton. They are then boosted to the lab frame. We assume that either the quark or the diquark will take the full recoil effect of the TMD gluon. Suppose the quark is recoiled, then the final-state remnant quark is
\begin{align}
k_q' = k_q-k,
\end{align}
with $k^+$ determined by the requiring $(k_q-k)^2=m_q^2$.

\begin{figure}
\centering
\includegraphics[width=\columnwidth]{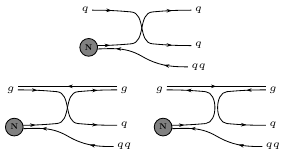}
\caption{(Color online) Color flow in the forward scattering of a quark (the first row) or a gluon (the second row) with a constituent quark of the nucleon in the large $N_c$ limit.}
\label{fig:color-flow}
\end{figure}

For partons generated in the shower algorithm, the colors of the final-state partons are already assigned. 
We implement the color exchange at the end of the shower: for each parton, one loops over its collision history and exchanges color with the TMD gluon.
The color of the remnant quark and diquark will be assigned accordingly to maintain the color neutrality of the system. This is shown in Fig. \ref{fig:color-flow}. The first diagram depicts the color flow for the forward scattering of a quark with a constituent quark of the nucleon, which is then broken into a quark-diquark pair. 
For a gluon (the second and third diagrams), there are equal chances for it to exchange color or anti-color indices with the constituent of the nucleon.
Finally, the color-neutral system is hadronized via the Lund string fragmentation mechanism \cite{Sjostrand:2014zea}.

\begin{figure*}[ht!]
\centering
\includegraphics[width=\textwidth]{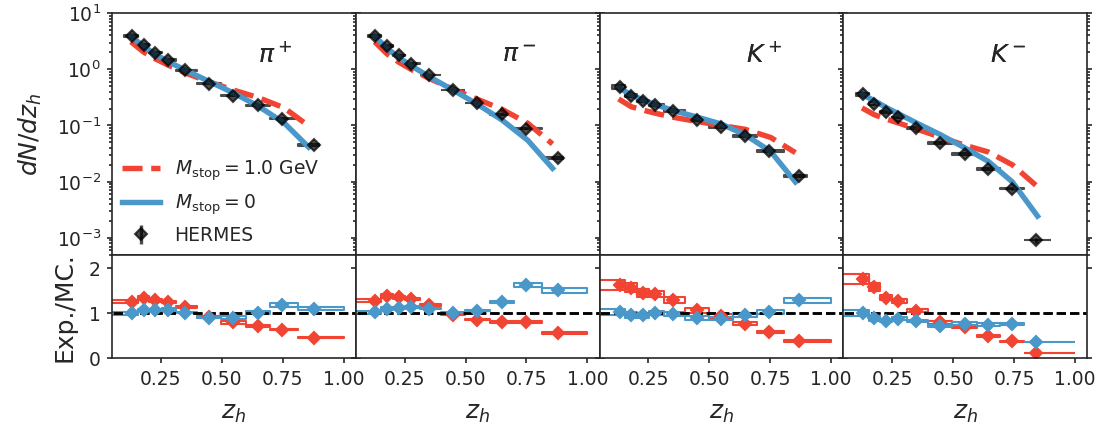}
\vskip-2em
\caption{(Color online) The single-inclusive spectra $dN/dz_h$ in $e+d$ collisions from Pythia8 compared to the HERMES experimental data~\cite{HERMES:2012uyd}. From left to right are the comparisons for $\pi^+$, $\pi^-$, $K^+$, and $K^-$. The ratio is shown in the lower panel. Simulations with the default Pythia8 parameters are shown in red. Simulations with $M_{\rm stop}=0$ GeV are shown in blue.}
\label{fig:ed-dNdz}
\end{figure*}
\begin{figure*}[ht!]
\centering
\includegraphics[width=\textwidth]{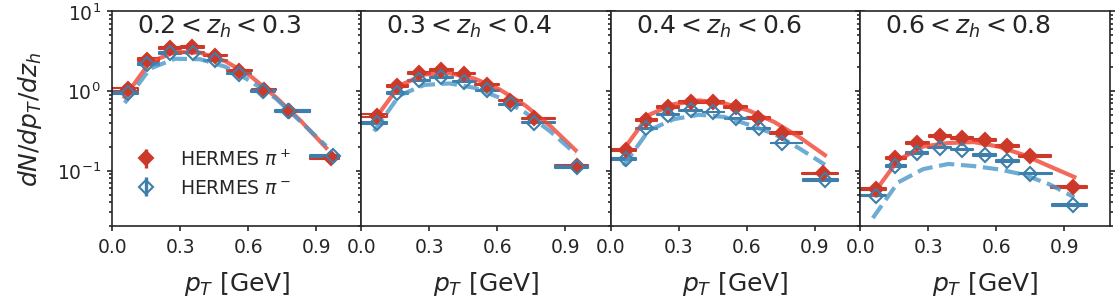}
\vskip-1em
\caption{(Color online) Transverse momentum spectra of pion from Pythia8 with $M_{\rm stop}=0$ GeV compared to HERMES experimental data \cite{HERMES:2012uyd} in different $z_h$ ranges. Red and blue lines and symbols are for $\pi^+$ and $\pi^-$ respectively.}
\label{fig:ed-dNdzdpT-pi}
\end{figure*}
\begin{figure*}[ht!]
\includegraphics[width=\textwidth]{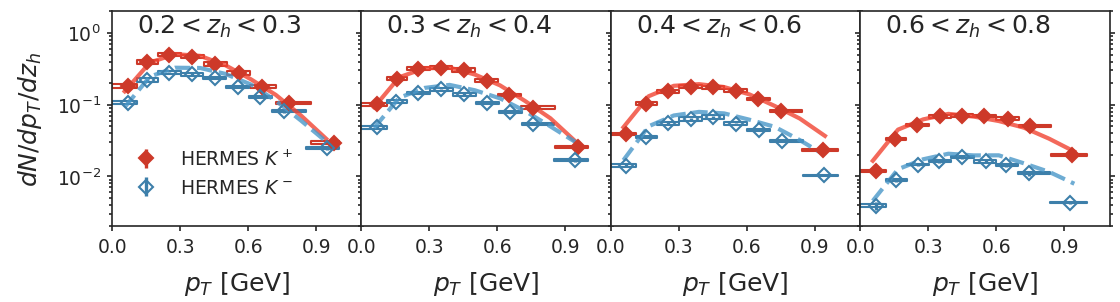}
\vskip-1em
\caption{(Color online) The same  as Fig.~\ref{fig:ed-dNdzdpT-pi}, but for kaons. Simulations are compared to HERMES experimental data \cite{HERMES:2012uyd}.}
\label{fig:ed-dNdzdpT-K}
\end{figure*}

\section{Hadron production in lepton-nucleus collisions }
\label{sec:results}

We apply eHIJING to study the single hadron production in the semi-inclusive deep inelastic scattering (SIDIS) process.
In the Breit frame, one can perform a three-dimensional (transverse and longitudinal momentum) study of the hadron production. 
The SIDIS cross-section normalized by the inclusive cross-section 
\begin{align}
\frac{dN_{e+A\rightarrow h}}{dz_h dp_T^2} = \frac{d\sigma_{e+A\rightarrow h}/dx_B/dQ^2/dz_h/dp_T^2 }{ d\sigma_{eA}(x_B, Q^2)/dx_B/dQ^2}
\label{eq:normalized_sidis}
\end{align}
provide a multidimensional calibration of the in-medium jet dynamics with respect to $x_B$, $Q^2$, $A$, $z_h$, $p_T$.
Here, $z_h = E_h/\nu$ is the energy fraction of the hadron relative to the energy of the virtual photon $\nu$ in the nuclear rest frame, and $p_T$ is the transverse momentum in the Breit frame.
We first study hadron production in $e$+$p$, where all the medium effects are turned off. Such simulations are performed in Pythia8, and we shall comment on some changes we made as compared to the default Pythia8 settings for $e$+$p$ simulations.
Then, eHIJING $e$+$A$ results are systematically compared to available data from the CLAS experiment at Jefferson Laboratory (JLab), the HERMES experiment at DESY, and the EMC experiment at CERN.

Another application of eHIJING is the study of di-hadron correlation.
This observable was proposed to disentangle the hadronic versus partonic energy loss picture. 
The original idea is that if two hadrons are produced from the same parton that loses energy in the medium, then a rescaled di-hadron correlation function should be similar to that in $e$+$p$ collisions.
On the contrary, if hadrons are formed inside the medium, independent energy loss or absorption of the two hadrons should strongly modify the correlation function.
We will make a realistic simulation of this observable using eHIJING.

Finally, we also make projections for experiments at the future Electron-Ion Collider (EIC) and Electron-Ion Collider in China (EicC). These new experiments cover a wide range of $x_B$ and $Q^2$ and we use eHIJING to simulate its capabilities to determine the $x_B, Q^2$ of the jet transport parameter.

\subsection{Baseline: SIDIS in electron-deuteron collisions}
\label{sec:results:baseline}
Compared to the default Pythia8 settings for $e$+$p$ DIS mode, the first change is that we used the global recoil instead of the dipole recoil scheme. The reason for this non-standard choice, explained in Sec. \ref{sec:default-vs-dipole}, is that, at the moment, we can only treat medium-modified showers as a pure final-state effect. 
The impact on observables are also discussed Sec. \ref{sec:default-vs-dipole}.
Furthermore, to compare to HERMES or CLAS data at fairly low $Q\gtrsim 1$ GeV, we decrease the minimum phase-space cut for hard processes of Pythia8 down to 1 GeV.

The results for the $z_h$-differential spectra of $\pi^{\pm}$ and $K^{\pm}$ in $e+d$ collisions are shown in Fig.~\ref{fig:ed-dNdz}.
Red lines are results simulated with the default hadronization settings of Pythia8 with data points in black. 
Ratios of simulated results over data  are shown in the bottom panel.

We find that the $z_h$ spectra are harder than those observed in the data.
To improve the model simulations, we further change the parameter ``$M_{\rm stop}$'' in the Pythia8 hadronization module. It plays a role in setting the minimum invariant mass $W_{\rm min}$ below which the standard string breaking stops. $W_{\rm min} = m_q + m_{\bar{q}} + M_{\rm stop}$ is the sum of $M_{\rm stop}$ and the constituent masses of the quark and anti-quark at the endpoints of the string.
We decrease $M_{\rm stop}$ from the default value $1$ GeV to zero, allowing strings to continue to break into softer hadrons. 
Simulations with $M_{\rm stop}=0$ are shown in blue in Fig. \ref{fig:ed-dNdz}.
The $z_h$ spectra are softened and the agreement with data is improved, especially for $K^\pm$.
For all simulations hereafter, we use $M_{\rm stop}=0$ in eHIJNG.

In Figs.~\ref{fig:ed-dNdzdpT-pi} and \ref{fig:ed-dNdzdpT-K}, the $p_T$ spectra of $\pi^\pm$ and $K^\pm$ are compared with the HERMES data. 
Each panel corresponds to the $p_T$ spectra in a different range of $z_h$. 
Simulation agrees well with the data in the range $p_T<1$ GeV, except for $pi^-$ spectra in the rightmost panel ($0.6<z_h<0.8$). This is because the $dN/dz$ of $\pi^-$ is already underestimated in the large $z_h$ region.  
In Pythia8, the transverse momentum of the hadron relative to the direction of the virtual photon comes from several sources:
\begin{itemize}
    \item The primordial $k_T$ of the quark inside the deuteron.
    They are parametrized as a Gaussian distribution with a $Q$-dependent width parameter \cite{Sjostrand:2014zea}.
    \item Momentum recoil of the leading quark from vacuum gluon emissions in the initial state parton shower. However, this is not a big effect in simulations for HERMES and CLAS experiments due to the limited phase space.
    \item Transverse momentum obtained during the string fragmentation. Again, this is parametrized as a Gaussian distribution with width $\sqrt{\langle k_T^2\rangle}  = 0.335$ GeV.
\end{itemize}
With these results, we consider the non-perturbative models in Pythia8 to provide a good description of the semi-inclusive hadron production $e+d$ baseline for $p_T<1$ GeV.
This is sufficient for studying most of the data at CLAS, HERMES and EMC.
However, at higher colliding energies such as $\sqrt{s} = 300$ GeV in H1 experiment at HERA, higher-order hard matrix elements are imperative to understand the large $p_T$ region. 
%Improvements by including NLO event generation, such as the SHERPA event generator \cite{Gleisberg:2008ta}, are possible in the future. 

\begin{figure*}[ht!]
    \centering
    \includegraphics[width=\textwidth]{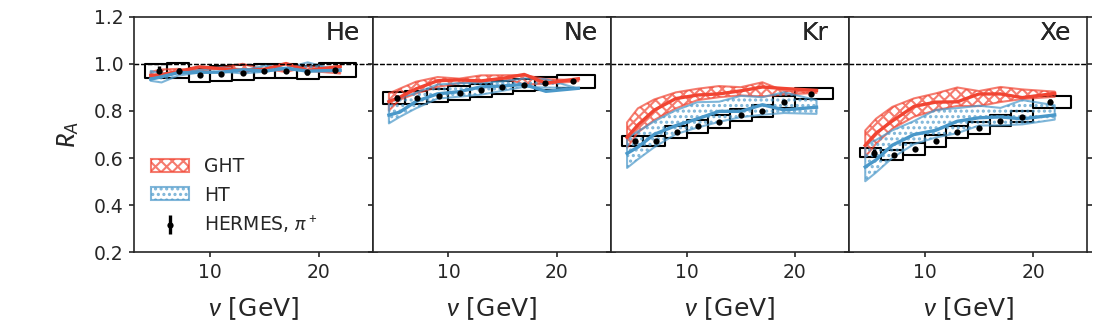}
    \caption{(Color online) Ratio of $\pi^+$ multiplicities between $e+A$ and $e+d$ collisions $R_A$ as a function of $\nu$. Blue dotted bands are simulations using the higher-twist formula, and red crossed bands are results using the generalized higher-twist formula. The central lines, upper and lower bounds of the bands correspond to the variation of the scale factor $K$ as in Fig.~\ref{fig:Qs}. Simulations are compared to HERMES experimental data \cite{HERMES:2007plz}.}
    \label{fig:Rnu_pi+}
\end{figure*}
\begin{figure*}[ht!]
    \centering
    \includegraphics[width=\textwidth]{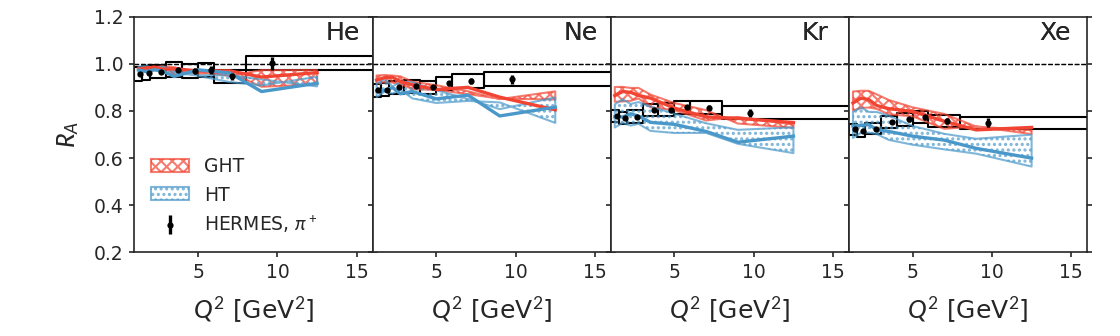}
    \caption{ (Color online) The same as Fig.~\ref{fig:Rnu_pi+}, but for the $Q^2$ dependence of the nuclear modification $R_A$. Simulations are compared to HERMES experimental data \cite{HERMES:2007plz}.}
    \label{fig:RQ2_pi+}
\end{figure*}

\subsection{Nuclear modification single-hadron production}
\label{sec:results:eA-FF}

Nuclear modifications to semi-inclusive hadron production have been studied in the following experiments:
\begin{itemize}
\item The first observation is made by the EMC experiment \cite{EuropeanMuon:1991jmx} in the charged hadron spectra with $d$, ${}^{12}$C, ${}^{64}$Cu, and ${}^{120}$Sn targets. The muon beam energy is $E =100$, $280$ GeV and the center of mass energy $\sqrt{s_{\mu N}} = 13.7$, $22.9$ GeV.
\item The HERMES experiment 
\cite{HERMES:2007plz} bombards electron/position on fixed targets of $d$, ${}^{3}$He, ${}^{14}$N, ${}^{20}$Ne, ${}^{84}$Kr, and ${}^{131}$Xe. The electron beam energy is $E = 27.6$ GeV and $\sqrt{s_{eN}} = 7.24$ GeV. Compared to EMC experiment, HERMES has a differential dataset and provides particle identification.
\item Finally the CLAS experiment \cite{CLAS:2021jhm} has various targets including $d$, ${}^{12}$C, ${}^{56}$Fe, and ${}^{208}$Pb. The electron beam energy is $E = 5.014$ GeV and $\sqrt{s_{eN}} = 3.21$ GeV. Nuclear experiments with 12 GeV electron beam energy are expected in the future.
\end{itemize}

These experiments probe the region $x_B\gtrsim 0.1$, where the hard processes are dominated by the LO quark scattering. This corresponds to the geometric picture shown on the left of Fig. \ref{fig:spacetime}, which will be used to understand the simulation and data.

The average $Q^2$ for CLAS and HERMES experiments are about 1-3 GeV${}^2$, while the dynamically generated scale in the medium is about $\nu/L$ is of order $1$ GeV${}^2$. The parton energy loss in such a kinematic region is large and thus ideal for testing the medium modifications in eHIJING. 
$\nu$ ranges from a few to about 20 GeV. The formation time of hadrons produced at small $z_h$ may be formed inside the nucleus $\tau_h \sim z_h\nu/\Lambda_{\rm QCD}^2 < L$, and then undergo hadronic interactions that are not included in eHIJING. 
For the EMC experiment, the average $Q^2$ reaches $10.2$ to $12.3$ GeV${}^2$, which is then used to constrain the virtuality evolution of the patron shower in eHIJING. The average $\nu$ exceeds 50 GeV, and we would expect the hadronic interactions to be negligible for a large range of $z_h$.

The medium effects are presented as the so-called nuclear modification factor, which is the ratio of the normalized SIDIS cross section between $e$+$A$ and $e$+$d$ collisions (or $e$+$p$ for EIC and EicC),
\begin{align}
  R_A^h(\nu,Q^2; z_h, p_T) = \frac{dN_{e+A\rightarrow h}/dz_h/dp_T^2}{dN_{\{ed,ep\}\rightarrow h}/dz_h /dp_T^2}.
\end{align}
$dN/dz_h /dp_T^2$ has been defined in Eq. (\ref{eq:normalized_sidis}).
Events in CLAS and HERMES experiments are selected with $Q^2>1$ GeV${}^2$, photon-nucleon center-of-mass energy $W_{\gamma N}>2$ GeV, and the inelasticity $y = (p\cdot q)/(p\cdot l_1) < 0.85$. Events in the EMC experiment is required to have $Q^2>2$ GeV${}^2$ and $y<0.85$. Additional cuts on $x_B$, $\nu$ will be specified later in the discussion.
Furthermore, the HERMES experiment only counts hadrons in the photon-going direction in the photon-nucleon center-of-mass frame in order to suppress hadrons from target fragmentation. These kinematic constraints are imposed in simulations.

\subsubsection{$\nu$ and $Q^2$ dependence of hadron production}
$R_A$ for $\pi^+$ ($z_h>0.2$) as a function of $\nu$ is shown in Fig. \ref{fig:Rnu_pi+} for different nuclear targets, from small (left) to large (right) mass number.
The suppression is stronger in larger nuclei and for jets with smaller $\nu$. 
The red hatched bands and the blue dotted bands are simulations using either the generalized higher-twist formula (GHT) or the higher-twist formula (HT). The parameters for the TMD distribution of nuclear gluons are the same for the two choices. 
The bands show the variation of the $K$ parameter from $2$ to $8$ in the TMD gluon distribution model (see Eq.~(\ref{eq:phig}) ) and the solid lines correspond to the set with $K=4$.
With the same $K$ factor, HT approach results in stronger nuclear modification than the GHT approach.
This is not surprising, as can be explained using Fig.~\ref{fig:compare_HT_Gen_2}: with the same input to the TMD gluon distribution,  medium corrections in the GHT approach is weaker than the HT approach.
For the rest of this section, we will continue to use the same set of $K$ parameters and this difference between GHT and HT approaches persists.

With the current range of the $K$ parameter, the HT simulation gives a ``better'' description of the data.
However, we remark that here the variation of $K$ is only intended to show the sensitivity of $R_A$ to the magnitude of the jet-medium interaction rates. The values of $K$ have not been tuned to data. So Fig. \ref{fig:Rnu_pi+} does not mean that the GHT approach is less effective than the HT approach.
If one tunes $K$ independently for the HT and the GHT approach to fit the data. Then, the HT approach would require a smaller $K$, thus a smaller jet transport parameter than the GHT approach.

Interestingly, such a difference has been seen in previous studies. The higher-twist study in Ref.~\cite{Chang:2014fba} suggests a smaller jet transport parameter than those values used in a study based on soft-collinear-effective-theory with Glauber gluons \cite{Li:2020zbk}, which, if one applies the soft gluon approximation, reduces to the GHT formula with Gyulassy-Wang model in Eq. (\ref{eq:GWmodel}).
In an examination of the connection between the generalized higher-twist and higher-twist approach to radiative parton energy loss \cite{Cao:2020wlm}, it is also realized that the effective radiative jet transport parameter in the higher-twist approach is smaller than that in the generalized higher-twist approach by a logarithmic factor of $2\ln(Q^2/Q_s^2)$.

In Fig.~\ref{fig:RQ2_pi+}, we plot the $Q^2$ dependence of $R_A^{\pi^+}$, integrated over $z_h>0.2$ and $\nu>6$ GeV.
The measured $R_A$ is almost independent of $Q^2$ for $1<Q^2<10$ GeV${}^2$, while the simulated $R_A(Q^2)$ slightly decreases with increasing $Q^2$ but is consistent with the data within experimental errors.

\begin{figure*}[ht!]
    \centering
    \includegraphics[width=.75\textwidth]{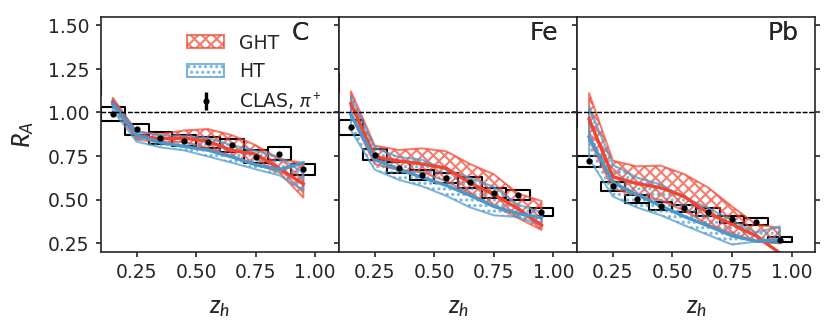}
    \caption{(Color online) Nuclear modification of the fragmentation function of $\pi^+$ as a function of $z_h$, $R_A=D_{eA}(z_h)/D_{ed}(z_h)$. Simulations with higher-twist formula (blue dotted bands) and generalized formula (red crossed bands) are compared to the measurements from the CLAS experiment \cite{CLAS:2021jhm} with an electron beam energy $E_e=5.014$ GeV.}
    \label{fig:Rz_pi+_CLAS}
\end{figure*}
\begin{figure*}[ht!]
    \centering
    \includegraphics[width=\textwidth]{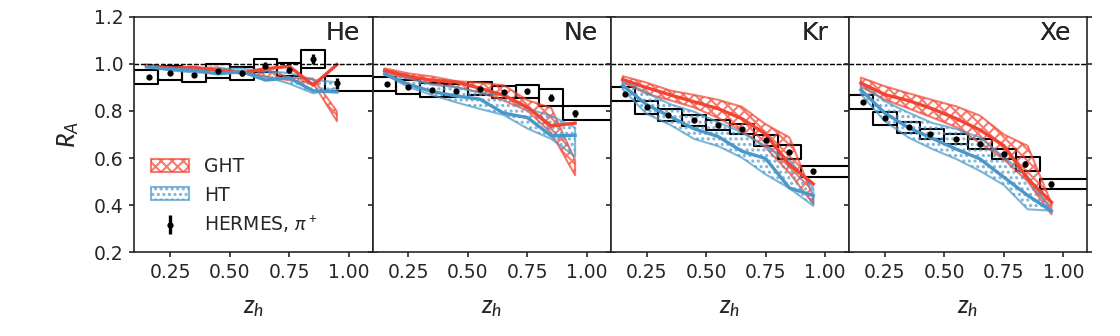}
    \caption{ (Color online) The same as Fig.~\ref{fig:Rz_pi+_CLAS}, but compare to the HERMES data \cite{HERMES:2007plz} with an electron beam energy $E_e=27.6$ GeV.}
    \label{fig:Rz_pi+_HERMES}
\end{figure*}
\begin{figure*}[ht!]
    \centering
    \includegraphics[width=.75\textwidth]{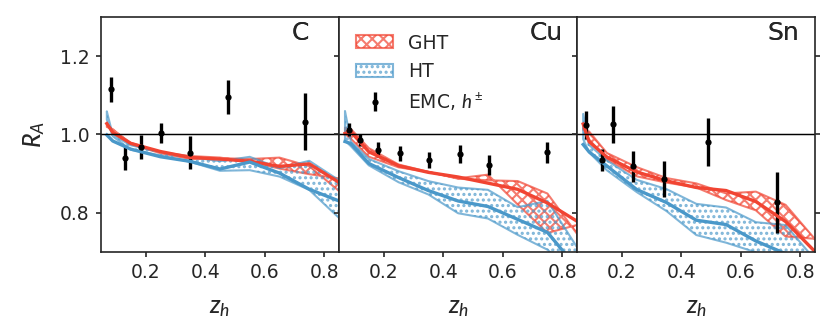}
    \caption{(Color online) Similar to Fig.~\ref{fig:Rz_pi+_CLAS}, but compare to  $R_A$ from the EMC data \cite{EuropeanMuon:1991jmx} on charged particles with electron beam energies $E_e=100,280$ GeV.}
    \label{fig:Rz_pi+_EMC}
\end{figure*}

\subsubsection{$z_h$ dependence of the modification} 
$R_{A}$ for $\pi^+$ as a function of $z_h$ from eHIJING is compared to CLAS in Fig.~\ref{fig:Rz_pi+_CLAS}, HERMES in  Fig.~\ref{fig:Rz_pi+_HERMES} and EMC data in Fig.~\ref{fig:Rz_pi+_EMC}, respectively.
The general trend across all three experiments is that the $z_h$ spectra are suppressed at large $z_h$. This is understood as a result of parton energy loss in the nuclear matter and is nicely described by both GHT and HT-based simulations. 
Because the fragmentation function sharply falls off to zero as $z_h$ approaches one, a small amount of energy loss of the quark can cause a drastic reduction of the produced hadrons at large $z_h$. This explains why the suppression is the strongest when $z_h\approx 1$.

Hadrons produced from the nuclear remnant contribute to small $z_h$, which complicates the interpretation of the observed modification. For example, the CLAS experimental data display an enhancement at small $z_h<0.1$ (not included in the plotting range of Fig. \ref{fig:Rz_pi+_CLAS}). 
To suppress remnant, the HERMES experiment  only uses particles produced in the photon-going side in the photo-nucleus center of mass frame ($x_F>0$). 
%The $R_{A}$ of the does not increase as fast as those in the CLAS data at small $z_h$.
In the EMC experimental data, the target fragmentation region is located at much smaller $z_h$, outside the kinematic cuts. 

The three experiments cover a wide range of mass numbers from $A=4$ to $A=208$. The $A$ dependence of $R_A$ is also well reproduced by eHIJING. Again, with the same input, simulation with the GHT approach results in a smaller suppression than the HT approach. 

In the future, we will also consider the inclusion of heavy-flavor quark energy loss in the eHIJING framework. 
The advantage of the heavy-flavor probe is that the heavy meson fragmentation function has a peak in $z_h$. In comparison, the pion fragmentation function is relatively featureless.
As a result, modification of the $z_h$ spectra of heavy meson exhibits a non-monotonic structure that is more sensitive to the spectral shift caused by quark energy loss~\cite{Li:2020zbk}.

Regarding the beam-energy dependence of $R_{A}(z_h)$, the model seems to under-estimate the suppression in low-energy collisions in CLAS experiment but over-predicts the suppression in EMC. 
On the one hand, part of the discrepancy can be attributed to the absence of hadronic absorption which should be more important in lower energy collisions. On the other hand, since $x_g \sim Q_s^2/2M\nu$, it is also possible to change the $x$-dependence of the gluon distribution so that the jet transport parameter should increase slower with $\nu$ in high-energy collisions. 
A systematic calibration, incorporating both hadronic effects and a flexible model for the small-$x$ gluon distribution, can be performed in the future using existing data to make a reliable prediction for future EIC experiments.

\subsubsection{$p_T$ dependence of the modification}
\label{sec:pt-dep-mod}

\begin{figure*}
    \centering
    \includegraphics[width=\textwidth]{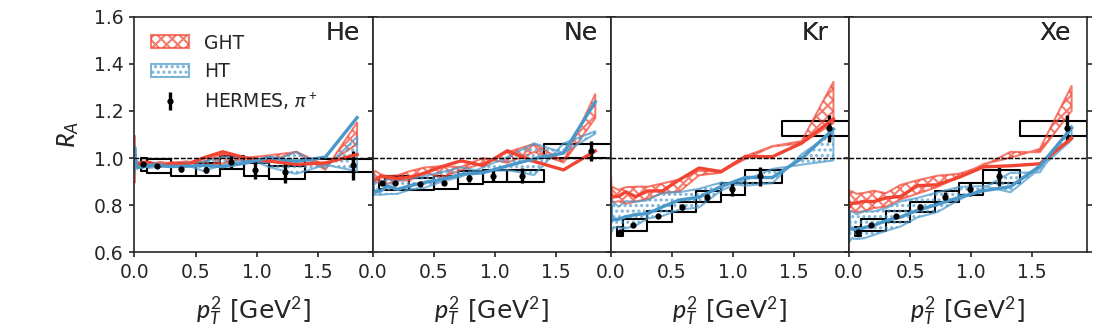}
    \caption{ (Color online) The modification $R_A$ of $\pi^+$ as a function of hadron transverse momentum squared from eHIJING with generalized higher-twist (red cross-hatched) and higher-twist (blue dot-hatched) approach as compared to HERMES data \cite{HERMES:2007plz}.}
    \label{fig:RpT_pi+}
\end{figure*}

Besides energy loss, the parton also undergoes transverse momentum broadening, which leads to the hardening of the shape of the $p_T$ spectrum of hadrons in $e$+$A$ collisions.
This can be seen in Fig. \ref{fig:RpT_pi+}, where the $R_A$ for $\pi^+$ increases with $p_T^2$.
Events are required to have $\nu>6$ GeV and $Q>1$ GeV and the spectra are integrated over $z_h>0.2$. The $A$ dependence of the slope of $R_A$ of simulations agrees with the data. Again, the GHT approach results in a weaker modification than the HT approach.

\begin{figure*}
    \centering
    \includegraphics[width=.75\textwidth]{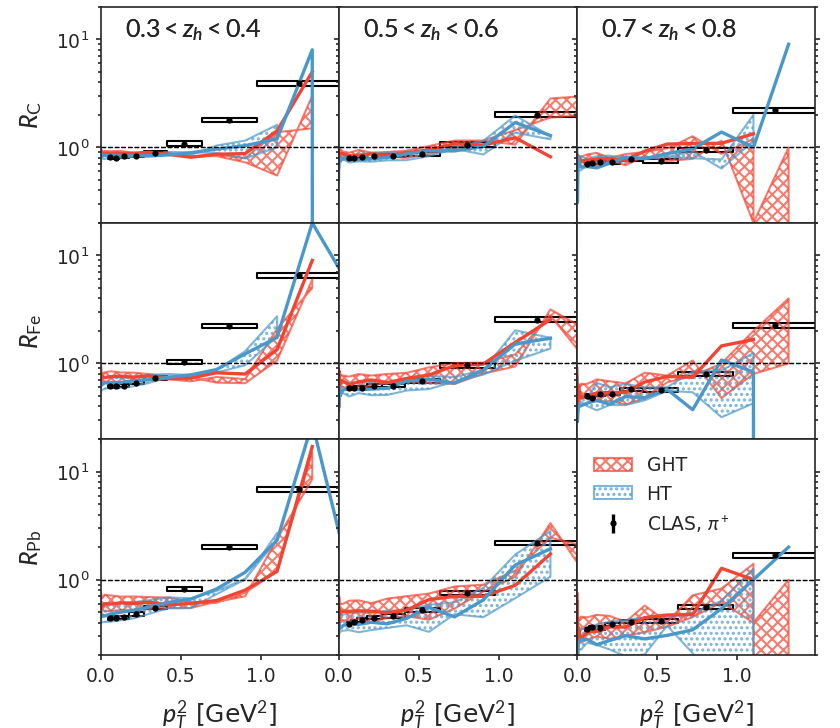}
    \caption{(Color online) Nuclear modification of the TMD fragmentation function $R_A(z_h, p_T) = D_{eA}(z_h,p_T)/D_{ed}(z_h, p_T)$ as a function of $p_T^2$ for different target nuclei (different rows) and different regions of $z_h$ (different panels) 
    from eHIJING with the generalized higher-twist (red cross-hatched) and higher-twist (blue dot-hatched) approach
    as compared to CLAS experimental data \cite{CLAS:2021jhm}.}
    \label{fig:CLAS:RzpT_pi}
\end{figure*}

\begin{figure*}
    \centering
    \includegraphics[width=.75\textwidth]{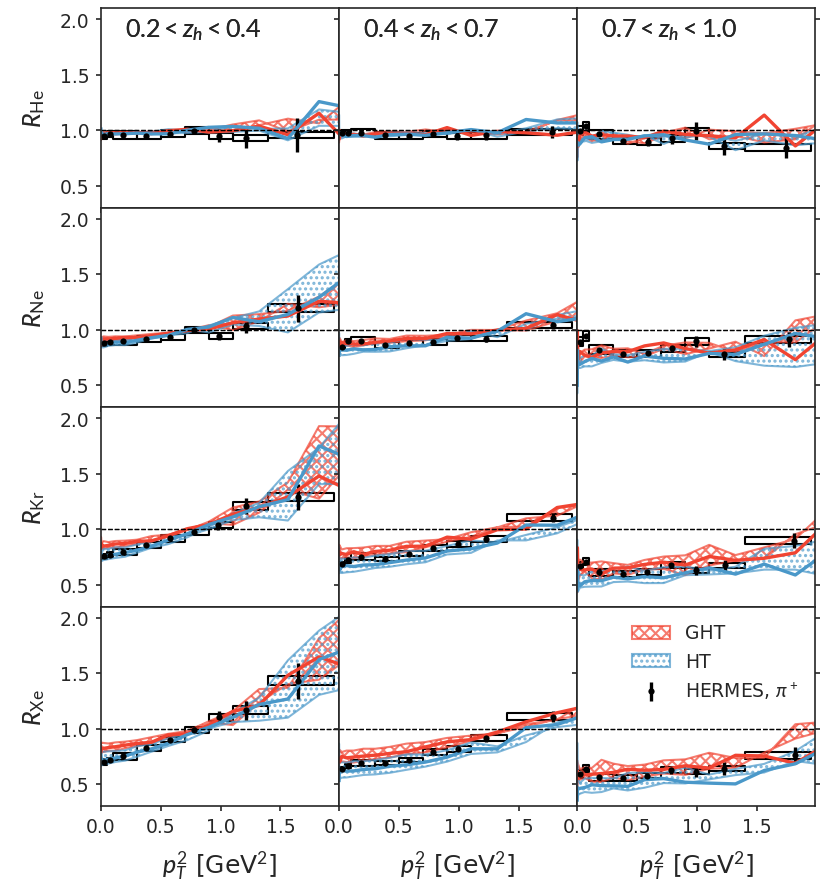}
    \caption{(Color online) Similar to Fig.~\ref{fig:CLAS:RzpT_pi}, but compare to HERMES experimental data \cite{HERMES:2007plz}.}
    \label{fig:HERMES:RzpT_pi}
\end{figure*}

To further elucidate the interplay between parton energy loss and transverse momentum broadening, we investigate the double differential modification $R_A(z_h, p_T^2)$.
Figs.~\ref{fig:CLAS:RzpT_pi} and \ref{fig:HERMES:RzpT_pi} compare the simulated $R_A(z_h, p_T)$ to CLAS and HERMES data, respectively. 
Each row shows the ratio as a function of $p_T^2$ for different targets; different columns vary the range of $z_h$.
The key to understanding this 2D observable is that a parton that undergoes more multiple collisions is also likely to lose more energy.
There are two reasons for such a correlation:
\begin{enumerate}
    \item The random fluctuation of the path length $L^+$ correlates with the average amount of momentum broadening ($\propto L^+$) and the average energy loss ($\propto (L^+)^2$).
    \item Even for a fixed path length, the way we construct the stochastic medium-modified splitting function introduces an additional correlation between momentum broadening and medium-induced radiative energy loss. Remember that the number of collisions follows a Poisson distribution around the average number of multiple collisions. When there are no collisions, the parton is unmodified and has zero radiative energy loss.
\end{enumerate}
This correlation leads to a survival bias, in the sense that hadrons remaining in the large $z_h$ region must, on average, acquire less momentum broadening.
Using the idea of survival basis, we can understand why the $p_T^2$ slope of $R_A$ decreases when $z_h$ increases, which is true for both CLAS and HERMES data.
This cannot be explained if one only implements collision-number and path-length averaged medium-modification splitting functions.
In Sec. \ref{sec:results:eA-dpT2}, ``survival bias'' will also help us to understand why the broadening in the variance of the transverse momentum spectra $\Delta p^T_2$ drops down to zero as $z_h$ approaches unity.


\subsubsection{The flavor dependence of $R_A$}

\begin{figure}
    \centering
    \includegraphics[width=\columnwidth]{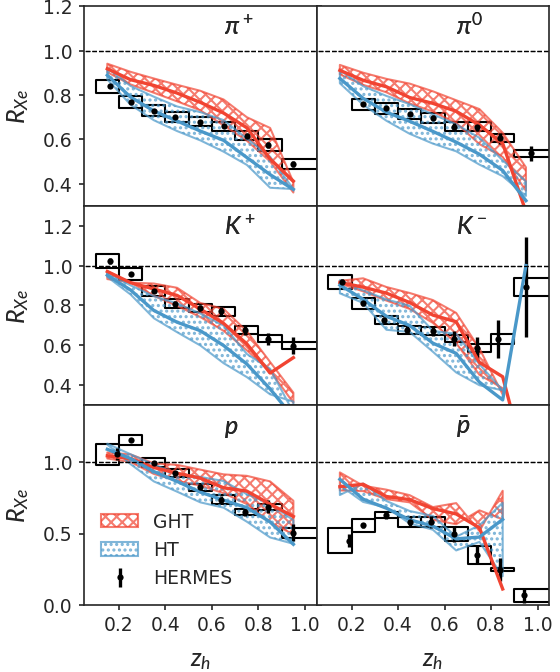}
    \caption{(Color online) The nuclear modification factor of fragmentation function $R_A(z_h)$ for different species of hadrons from eHIJING with generalized higher-twist (red cross-hatched) and higher-twist (blue dot-hatched) approach as compared to HERMES data \cite{HERMES:2007plz}. From left to right, top to bottom, $R_A$ of $\pi^+$, $\pi^0$, $K^+$, $K^-$, $p$, and $\bar{p}$ in $e$-Xe collisions are shown.}
    \label{fig:Rz_speices}
\end{figure}

Up to this point, we have only discussed the nuclear modification of $\pi^+$ for which the fragmentation functions in vacuum are well studied. 
Kaon and proton productions are complicated by in-medium strangeness production and flavor conversion \cite{Schafer:2007xh, Chang:2014lla}\footnote{When using GHT or HT formula under soft gluon approximations,  we have neglected medium-modifications to the $g\rightarrow q\bar{q}$ splitting functions ($q=u,d,s$). We will go beyond the soft gluon approximation in future versions of eHIJING.}, baryon production mechanism \cite{Kharzeev:1996sq}, as well as different hadronic absorption cross sections \footnote{The low-energy cross-section between $K^-$-$N$ and $K^+$-$N$ and between $p$-$N$ and $\bar{p}$-N are very different. This is not the case between $\pi^+$-$N$ and $\pi^-$-$N$ collisions}. 
Even though eHIJING does not include the aforementioned mechanisms, a systemic study of the flavor dependence of $R_A$ with pure nuclear PDF and medium-modified shower effects can still shed light on the problem. 
In Fig.~\ref{fig:Rz_speices}, we compare $R_A(z_h)$ for $\pi^+$, $\pi^0$, $K^+$, $K^-$, $p$, and $\bar{p}$. 
The $\pi^+$ modification (top left) has been discussed earlier. 
Using the same set of parameters as $\pi^+$, the description of neutral pions (top right) is also satisfactory. 
For kaons, data suggest that $K^-$ (middle right) is slightly more suppressed than $K^+$ (middle left) at intermediate $z_h$, which is not captured by the simulation. 
The explanation for this discrepancy must be rooted in the different valance structures of $K^+(u\bar{s})$ and $K^+(\bar{u}s)$ but can differ in details.
For example, at the partonic level, the $\bar{u}$ quark is more likely to be absorbed by the $u$-quark-rich nuclear matter and result in a stronger suppression of $K^-$ production.
At the hadronic level, it can be explained by a larger $K^-$-$N$ cross-section than $K^+$-$N$ at low energy, which may apply to low-$z_h$ hadrons that formed inside the nucleus. 
In either case, the $R_A^{K^+}$ versus $R_A^{K^-}$ separation may carry information on the valence structure of the heavy nuclei.
Finally, there is a more drastic difference between the $R_A$ of proton and anti-proton. This can be qualitatively understood with the same reasoning as for kaons.

Summarizing the comparison to nuclear modification factor $R_A$ of the semi-inclusive hadron production, we find the current model with multiple collisions and medium-induced gluon bremsstrahlung can describe the general trend of $\pi^+$ modification as a function of $\nu, Q^2, z_h$, and $p_T$. 
With the same input to the nuclear gluon TMD, the generalized higher-twist approach results in smaller medium modifications than the higher-twist approach.
Finally, the current modeling cannot fully describe the different suppression patterns between $K^+$ v.s. $K^-$ and $p$ v.s. $\bar{p}$. This requires a more detailed modeling of the interaction between the partons/hadrons and the valence content of the nucleus.

\subsection{Transverse momentum broadening}
\label{sec:results:eA-dpT2}

\subsubsection{A theoretical estimation}
It is proposed that the difference of the variance of the hadron transverse momentum distribution in $e+A$ and $e+d$ collisions offers more direct access to the value of $\hat{q}$~\cite{HERMES:2009uge},
\begin{align}
   \langle\Delta p_T^2\rangle \equiv 
  \langle\Delta p_T^2\rangle_{eA} - \langle\Delta p_T^2\rangle_{ed}.
\end{align}
At leading order, $\langle\Delta p_T^2\rangle$ has been calculated in the higher-twist approach \cite{Guo:1998rd}, and the momentum broadening is proportional to the jet transport parameter times the path length $L$ of the nuclear matter.
With radiative correction, it is shown that the soft gluon emission effect can alter the $L$ dependence of the $\langle\Delta p_T^2\rangle \propto L^{1+2\sqrt{\alpha_s 
C_A/\pi}}$ at large path length \cite{Blaizot:2014bha}.
Ref. \cite{Ru:2019qvz} uses the NLO higher-twist formula, includes the effect of fragmentation in the calculation of $p_T$ broadening of hadrons $\Delta \langle p_T^2\rangle$, and performs a global extraction of the jet transport parameter in the cold nuclear matter. 
In computing the hadron transverse momentum broadening, Ref. \cite{Ru:2019qvz} uses an approximate extension to the collinear fragmentation function. The hadron carries approximately $z_h$ fraction of the quark's transverse momentum\footnote{Gluon fragmenting to hadron is neglected in the LO estimation.}, so the $p_T$ broadening for the hadron is,
\begin{align}
\langle\Delta p_T^2\rangle \approx z_h^2 \langle\hat{q}_F L\rangle = z_h^2\frac{C_F}{C_A}Q_s^2(x_B, Q^2),
\label{eq:LO_dpT2}
\end{align}
where we have replaced $\langle\hat{q}_FL\rangle$ by the average value of the saturation scale $Q_s^2(x_B, Q^2)$ times the quark over gluon ratio of color Casimir factors. 

\begin{figure*}
    \centering
    \includegraphics[width=\textwidth]{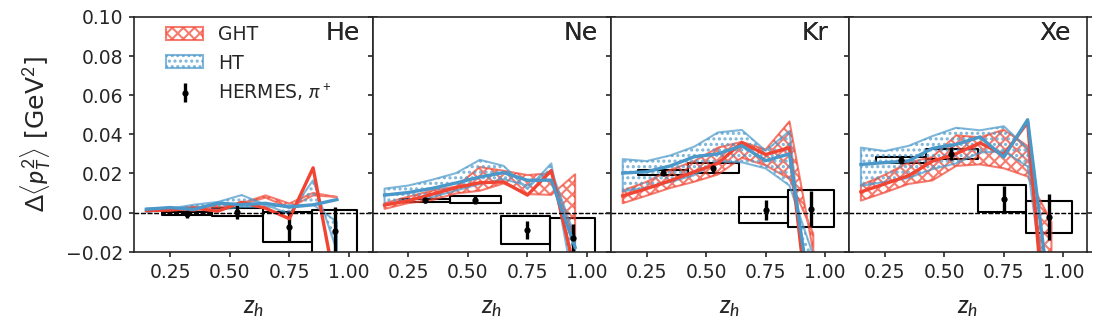}
    \caption{(Color online) Transverse momentum broadening $\Delta\langle p_T^2\rangle$ of $\pi^+$ as a function of $z_h$ from eHIJING with generalized higher-twist  (red cross-hatched) and higher-twist (blue dot-hatched) approach as compared to HERMES data \cite{HERMES:2009uge}.}
    \label{fig:DpTz}
\end{figure*}

\begin{figure}
    \centering
    \includegraphics[width=.8\columnwidth]{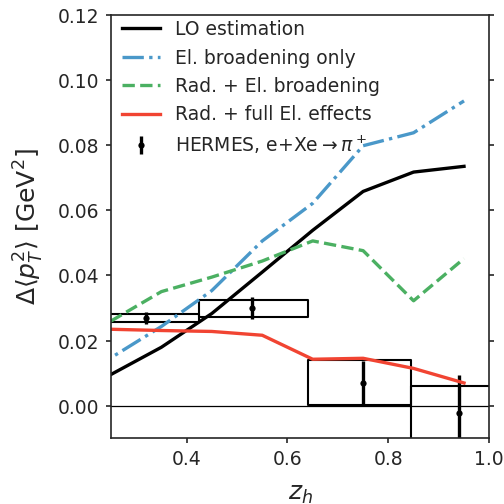}
    \caption{(Color online) Effect of momentum broadening, medium-modified evolution, and collisional energy loss on the shape of $\Delta\langle p_T^2\rangle(z_h)$ in $e$-Xe collisions. The black line shows the LO formula in Eq.~(\ref{eq:LO_dpT2}). The blue dash-dotted line is the eHIJING result with only elastic broadening. The green dashed line shows the combined effect of elastic broadening and medium-modified radiative correction. The red line includes the effect of the collisional energy loss. Data are from HERMES experiment \cite{HERMES:2009uge}.}
    \label{fig:DpTz-decomposed}
\end{figure}

\subsubsection{$z_h$ dependence of $\langle \Delta p_T^2\rangle$}

In Fig.~\ref{fig:DpTz}, the transverse momentum broadening of $\pi^+$ is plotted as a function of $z_h$ for different target nuclei.
For collision with a small nucleus like ${}^4$He, the $\langle\Delta p_T^2\rangle$ broadening is consistent with zero. For $Ne$, $Kr$, and $Xe$, both the simulated $\langle \Delta p_T^2\rangle$ and the data increase with $z_h$ first and then decrease to zero as $z_h$ approaches unity. As discussed in Sec.~\ref{sec:pt-dep-mod}, this non-monotonic feature is explained by a survival bias: 
the surviving hadrons at large $z_h$ mostly come from partons that suffer less scattering-induced energy loss\footnote{Both elastic collisions and induced gluon radiation contribute to the energy loss.} and therefore less transverse momentum broadening. 
For hadrons produced infinitely close to $z_h=1$, it cannot undergo any collisions with the nucleus, and there $\langle\Delta p_T^2\rangle (z_h\rightarrow 1)\rightarrow 0$.
The survival bias is absent in the leading-order formula in Eq.~(\ref{eq:LO_dpT2}), where the $p_T$ of the hadron is assumed to be $z_h$ fraction of the transverse momentum of parton.  
Consequently, the leading-order estimation yields a momentum broadening that always increases with $z_h$.

To further illustrate this interplay between the effect of parton energy loss and transverse momentum broadening in inclusive hadron spectra, we examine in Fig. \ref{fig:DpTz-decomposed} the relative importance of the elastic and radiative energy loss in the $p_T$ broadening of $\pi^+$ in $e$-Xe collisions in the large $z_h$ region.
The black line is obtained from the LO estimation using $Q_s^2(x_B, Q^2, T_A)$ and $R_A(z_h)$ obtained with the model using $K=4$. For such an estimation, we take the typical values of $x_B = 0.1$, $Q^2 = 2.5$ GeV${}^2$, $\langle T_A \rangle = 3r_0 A^{1/3}/4$. $R_A$ is simulated within the generalized higher-twist approach. As expected, the LO calculation monotonically increases with $z_h$. 
This is also confirmed in our simulation if we only include the effect of transverse momentum broadening while turning off the elastic energy loss and medium-induced radiation (the blue dash-dotted line). If one includes the induced radiation, the peak of the $z_h$ dependence of $\Delta\langle p_T^2\rangle(z_h)$ is shifted from $z_h=1$ towards lower $z_h$ (the green dashed line). However, it is still higher than the experimental data for $z_h>0.6$. 

Only with the inclusion of both elastic and radiative parton energy loss (red solid line), the agreement with data is much improved in the large-$z_h$ region. 
In the power counting of the jet sector, the small $x$ gluon in the nucleus has $k^+=0$, leading to vanishing collisional energy loss of the jet parton\footnote{This approximation is based on the fact that collisional energy is usually subleading to radiative energy loss at high energy}. 
But as we have seen in Sec. \ref{sec:model:remnant}, a negative $k^+=-\bfk^2/p_q^-\sim -Q_s^2/p_q^-$ is necessary for the simultaneous modeling of both the jet and the target remnant sectors and induces collisional energy loss of the jet parton.
The collisional energy loss fraction is of order $\Delta E / E \sim -Q_s^2/Q^2$.
We implement the collisional energy loss by retaining the finite $k^+$ in the jet sector, and the result is shown as the red solid line.
For the kinematics region used in Fig.~\ref{fig:DpTz-decomposed}, i.e., $Q>1$ GeV, it turns out that the collisional energy loss is very important to understand the $\Delta \langle p_T^2\rangle$ in the threshold region of $z_h\rightarrow 1$. Because $Q_s^2$ only increases logarithmically with $Q^^2$, one expects vanishing effects of elastic energy loss at large $Q^2$.

\subsubsection{$x_B$ and $Q^2$ dependence of $\langle \Delta p_T^2\rangle$}

\begin{figure*}
    \centering
    \includegraphics[width=\textwidth]{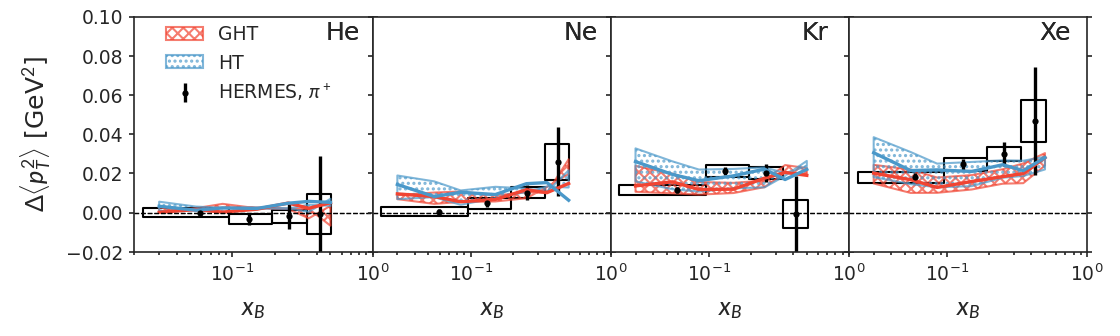}
    \caption{(Color online) Momentum broadening as a function of $x_B$ from eHIJING with generalized higher-twist (red cross-hatched) and higher-twist (blue dot-hatched) approach
    compared to HERMES \cite{HERMES:2009uge}  data.}
    \label{fig:DpT2xB}
\end{figure*}

\begin{figure*}
    \centering
    \includegraphics[width=\textwidth]{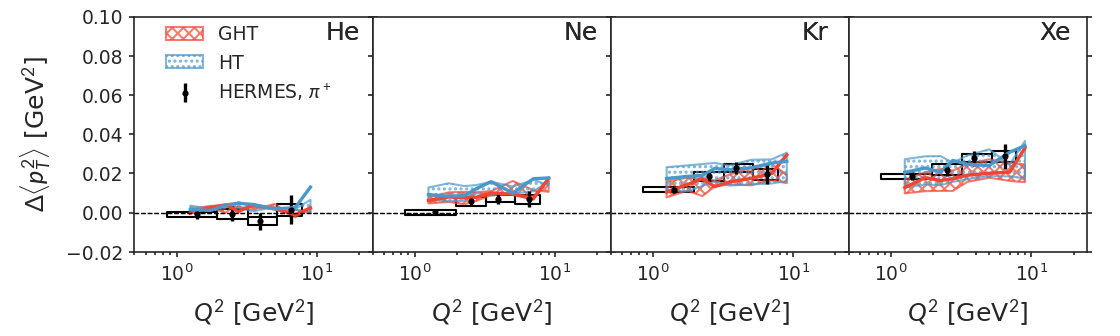}
    \caption{(Color online) Similar to Fig.~\ref{fig:DpT2xB} except as a function of the photon virtuality $Q^2$ compared to HERMES \cite{HERMES:2009uge} data.}
    \label{fig:DpT2Q2}
\end{figure*}

In Figs. \ref{fig:DpT2xB} and \ref{fig:DpT2Q2}, we integrate over final hadrons with $z_h>0.2$ and $W^2>10$ GeV${}^2$ and plot $\langle\Delta p_T^2\rangle$ as functions of $x_B$ and $Q^2$. The transverse momentum broadening increases with $x_B$ and the simulation is consistent with HERMES data in the large $x_B$ region. 
In the next section, we will predict that the transverse momentum broadening increases at small $x_B$ achievable at the future EIC, due to the $x_B$ dependence that we parametrized in $Q_s^2$. The $p_T$-broadening also increases logarithmically with $Q^2$.

\subsection{Medium-modified di-hadron fragmentation function}
\label{sec:results:eA-R2h}

Finally, we study the nuclear modification of the di-hadron fragmentation function to test the eHIJING model in describing more complicated observables.
The modification to the di-hadron fragmentation function is defined as the double ratio,
\begin{align}
R_{2h}(z_1, z_2) = \frac{\frac{1}{dN_{e+A\rightarrow h_1}/dz_1} \frac{d^2N_{e+A\rightarrow h_1 h_2}}{dz_1 dz_2}}{\frac{1}{d N_{ed\rightarrow h_1}/dz_1} \frac{d^2N_{ed\rightarrow h_1 h_2}}{dz_1 dz_2}}
\label{eq:R2h}
\end{align}
with $z_1>z_2$. In the numerator and denominator, the double hadron spectra are normalized by the single hadron spectra in $e+A$ and $e+d$ collisions, respectively.
The HERMES measurements select $z_1>0.5$ and $z_2<0.5$ hadron pairs from events with $W^2 > 10$ GeV${}^2$, $\nu>7$ GeV, $Q^2>1$ GeV${}^2$ and $y<0.85$.

The di-hadron correlation was initially proposed to distinguish between two different scenarios of hadron production in $e+A$ collisions~\cite{Majumder:2004pt,Majumder:2004wh}. One extreme situation (the LO parton picture) is that a hard parton loses energy in the medium and then fragments into hadrons in the vacuum; therefore, the di-hadron pair is produced from a common parton with a slightly reduced energy. If one assumes the di-hadron fragmentation does not strongly depend on the energy of the parton, the shape of the double ratio in Eq. (\ref{eq:R2h}) should not be strongly modified. 
Another extreme is that the two hadrons are formed very early in the medium (instantaneous hadron production), and one hadron interacts with the nucleus independently from the other. Because medium effects are stronger for the less energetic hadron, the shape of the double ratio will be modified. 
A more realistic situation is always in between the two extremes. 
Considering radiative correction to the parton picture~\cite{Majumder:2004pt,Majumder:2004wh}, there is a contribution where the two hadrons fragments from different daughter partons that have evolved independently in the medium. It modifies the shape of the double ratio in the NLO calculation.
On the other hand, the state-of-the-art hadronic transport model~\cite{Gallmeister:2007an,Buss:2011mx} implements a hadron formation time, before which the ``pre-hadrons'' interact with the medium with a reduced cross-section. 
As for di-hadron correlation from eHIIJNG, it follows the partonic picture modeling similarly to Refs.~\cite{Majumder:2004pt,Majumder:2004wh}; however, unlike the use of collinear fragmentation function in Refs.~\cite{Majumder:2004pt,Majumder:2004wh}, the hadronization of one parton is not completely independent of another parton in the Lund string model.

\begin{figure*}
    \centering
    \includegraphics[width=.8\textwidth]{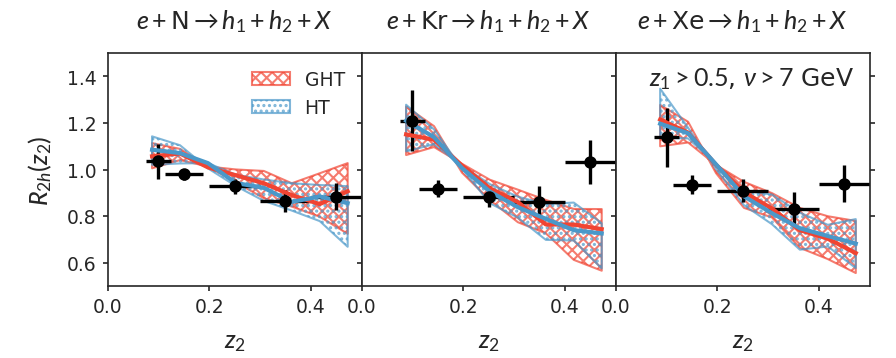}
    \caption{(Color online) Nuclear modification factor $R_{2h}(z_1, z_2)$ of the double hadron fragmentation function  as defined in Eq.~(\ref{eq:R2h}). Hadron pairs with opposite charges are excluded from the calculation. In each panel, $R_{2h}$ is plotted as a function of the momentum fraction of the subleading hadron $z_2$. Red cross-hatched and blue dot-hatched bands correspond to the generalized and higher-twist results, respectively, as compared to HERMES data \cite{HERMES:2005mar}.}
    \label{fig:dihadron}
\end{figure*}

In Fig.~\ref{fig:dihadron}, we compare the di-hadron nuclear modification factor from eHIJING to the HERMES data for N, Kr, and Xe targets from the left to the right panel.
The dihadron distribution function is already integrated for $z_1>0.5$ and is plotted as a function of $z_2$.
The nuclear modification factor is above unity at $z_2\approx 0.1$ and decreases at intermediate $z_2$ and, eventually, has the tendency of rising again when $z_2$ approaches 0.5. 
Simulation of eHIJING with either HT or GHT qualitatively describes the decreases of $R_{2h}(z_2)$ at small $z_2$, but fails to explain the rise near $z_2=0.5$.
The region $z_2\approx z_1\approx 0.5$ is an interesting kinematic region, the energy of the photon is carried almost exclusively by the two hadrons. We may need to better 
 understand the medium modification in this threshold region to address the discrepancy near $z_2\approx 0.5$.
The increase below $z_2= 0.2$ is underestimated, this may be due to the neglected hadronic interactions that are important for low-$z_h$ hadrons. 
An interesting observation is that the difference between the HT and GHT simulations is small.

\section{Predictions for future $e$+$A$ experiments}
\label{sec:eic}
In this last section, we use the eHIJING to make predictions for the CLAS experiment at the Jefferson Lab with 12 GeV electron beam energy, the Electron-Ion Collider at the Brookhaven National Laboratory (EIC) \cite{Accardi:2012qut}, and the proposed Electron-Ion Collider in China (EicC) \cite{Anderle:2021wcy}.

These future experiments can be performed with a variety of nuclear targets, higher luminosity, and large center-of-mass energy. It is also possible to do a precision study on the transport of partons in cold nuclear matter in the small-$x_B$ regions. The 12 GeV beam fixed target experiments at the Jefferson Lab have $\sqrt{s}\approx5.0$ GeV. The energy range of future EicC covers $10<\sqrt{s}<20$ GeV, and the future EIC covers a wide range of $20<\sqrt{s}<140$ GeV. The resulting hadron transverse momentum broadening at small $x_B$ provides a stringent test on the model for small-$x$ gluon distribution and parton dynamics.

\begin{table}[ht]
    \centering
    \begin{tabular}{c|ccc}
    \hline
    $Q^2$, $x_B$ & $(0.01,0.02)$ & $(0.1,0.2)$ & $(0.5,1)$\\
    \hline
    $(32,40)$ GeV${}^2$ & -  & C & - \\
    $(12,16)$ GeV${}^2$ & E & B & -  \\
    $(4,6)$ GeV${}^2$   & D & A & F \\
    \hline
    \end{tabular}
    \caption{(Color online) Ranges of $Q^2$ and $x_B$ for the regions labeled in Fig. \ref{fig:eic-kinematics}.}
    \label{tab:eic-kinematics}
\end{table}

\begin{figure}
    \centering
    \includegraphics[width=\columnwidth]{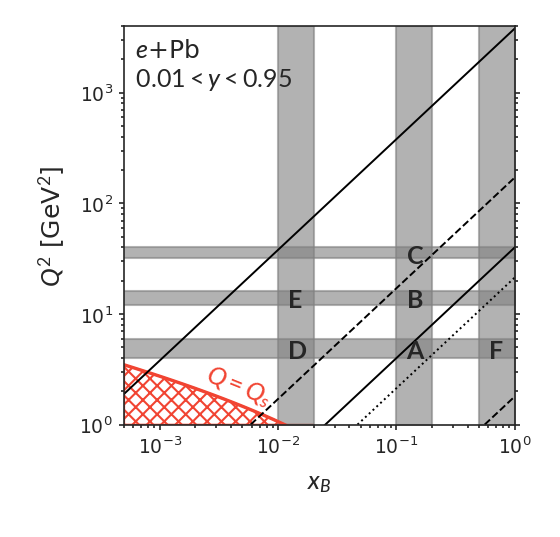}
    \caption{(Color online) The region enclosed between solid lines is the kinematic reach of a typical EIC setup ($E_e=10$ GeV, $E_N=100$ GeV) in the $x_B, Q^2$ plane. 
    The region enclosed between dashed lines is for EicC ($E_e=3$ GeV, $E_N=15$ GeV).  The region below the dotted line is accessible in the CLAS-12 fixed target experiment.
    The lower and upper bounds are estimated with $y_{\rm min}=0.01$ and $y_{\rm max}=0.95$.
    Single-inclusive observables are projected in regions A, B, C, D, E, and F. Note that we only selected regions where $Q^2\gg Q_s^2(x_B, Q^2)$ ensuring that $Q^2$ is always the hardest scale in the simulation.}
    \label{fig:eic-kinematics}
\end{figure}

In Fig. \ref{fig:eic-kinematics}, the accessible kinematic region for EIC ($E_e=10$ GeV, $E_N=100$ GeV), EicC ($E_e=3$ GeV, $E_N=15$ GeV) and CLAS-12 is shown between the solid lines, dashed lines and 
 below the dotted line, respectively. The upper and lower bounds are obtained for $0.01<y<0.95$.
Within the coverage of these experiments, we select six regions in the $(x_B, Q^2)$ plane, as listed in Tab. \ref{tab:eic-kinematics} to study the nuclear modification of hadron spectra. In particular, for simulations in $e$+Pb collisions, we have avoided the region where $Q^2<Q_s^2(x_B, Q^2)$ denoted as red shaded regions to ensure that $Q^2$ is the hardest scale in the problem.

\begin{figure*}
\centering
\includegraphics[width=.45\textwidth]{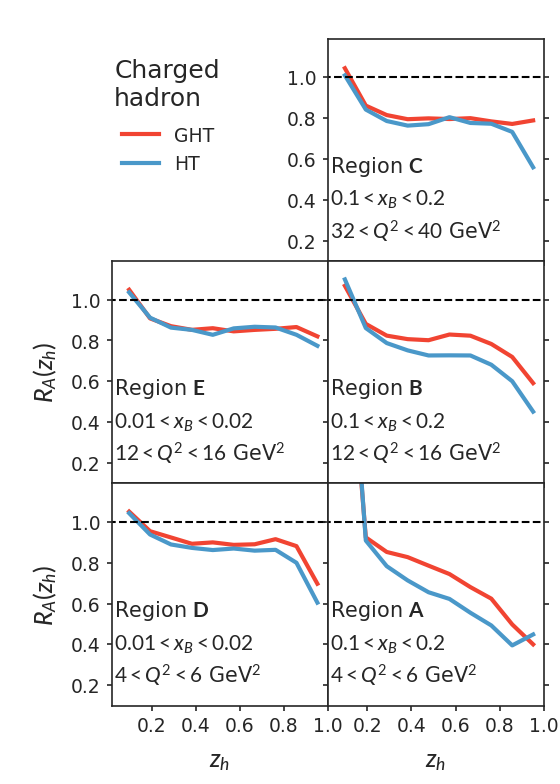}\includegraphics[width=.45\textwidth]{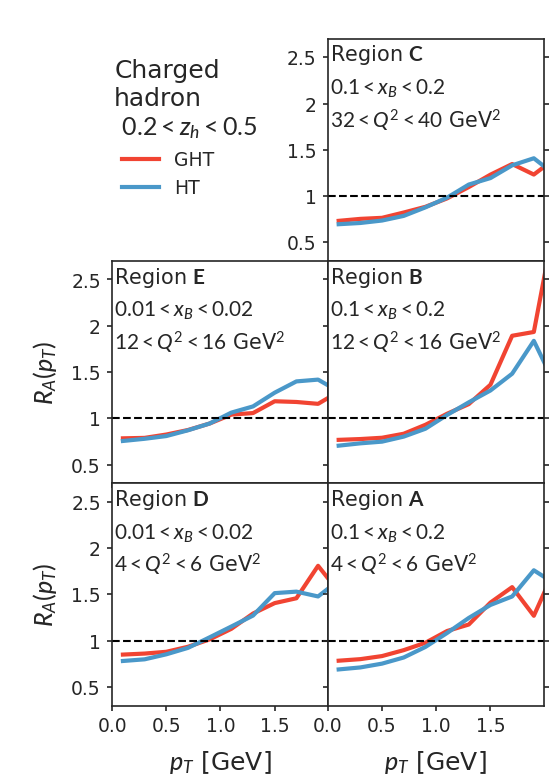} 
\caption{(Color online) The nuclear modification factor of the fragmentation function in regions as labeled in Fig. \ref{fig:eic-kinematics}. Left: modification of $D(z)$ for all charged hadrons. Right: modification of $D(p_T)$ for $0.2<z_h<0.5$. $K=4$ is used for the calculation.}
\label{fig:EIC-Dz-and-DpT2z}
\end{figure*}

In Fig. \ref{fig:EIC-Dz-and-DpT2z}, we show the nuclear medication factor of charged hadron fragmentation functions $R_A(z_h)$ (left) and $R_A(p_T)$ (right, with $0.2<z_h<0.5$) for regions A, B, C, D, E. The magnitude of the suppression factor $R_A(z_h)$ at large $z_h$ is reduced as one increases $Q^2$ or decreases $x_B$. This is because the parton energy in the nuclear rest frame $\nu$ increases for both cases and this effect overwhelms the increase of $\hat{q}$ at large $Q^2$ and small $x_B$.

\begin{figure*}
    \centering
    \includegraphics[width=.8\textwidth]{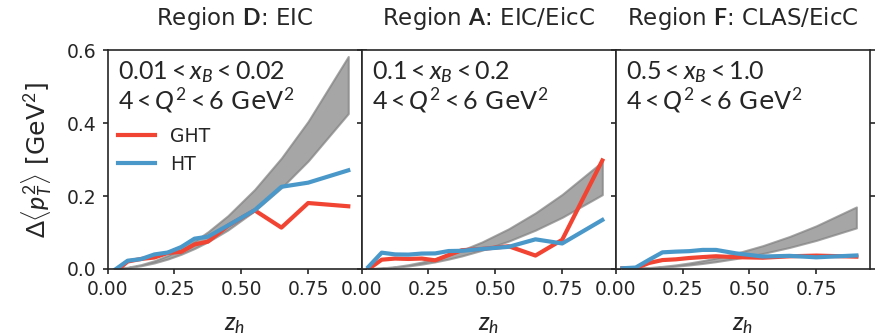}
    \caption{(Color online) The transverse momentum broadening of charged hadrons in DIS off nuclei as a function of the momentum fraction $z_h$ at different beam energies. Grey bands are LO results using the transport parameter evaluated for the indicated $x_B$ and $Q^2$ region and an averaged thickness for the Pb nuclei. The red and blue lines are from eHIJING with the higher-twist and generalized higher-twist approach, respectively.}
    \label{fig:EIC-DpT2z}
\end{figure*}

Shown in Fig.~\ref{fig:EIC-DpT2z} is the transverse momentum broadening as a function of hadron energy fraction $z_h$ in regions $d$, $A$, and $F$.
From the left panel to the right, the parton energy decreases, resulting in a smaller value of $\hat{q}$ and thus smaller momentum broadening. 
The LO formula works well in cases where the energy loss effect is small -- events with a large parton energy limit and hadrons away from the endpoint $z_h=1$.
At lower parton energy or when $z_h$ approaches unity, energy loss transports the broadened parton towards lower $z_h$ and effectively flattens the $\Delta \langle p_T^2\rangle(z_h)$.

\begin{figure}
    \centering
    \includegraphics[width=0.5\textwidth]{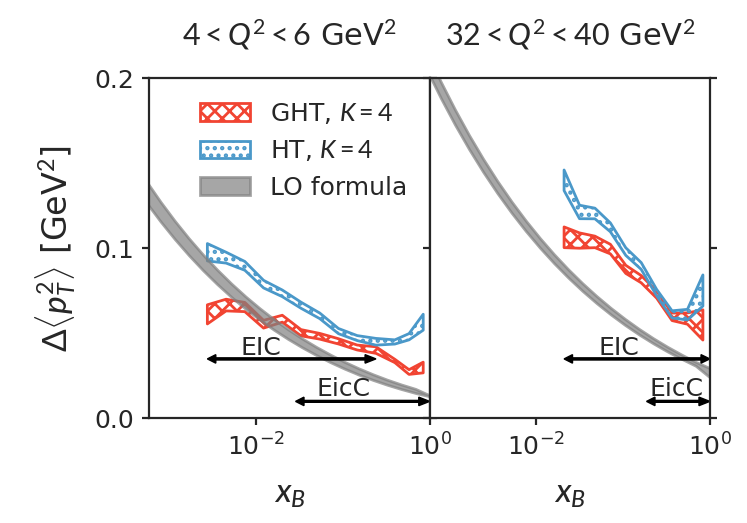}
    \caption{(Color online) The transverse momentum broadening of charged pions ($0.2<z_h<0.5$) in $e$+Pb as a function of $x_B$ for different $Q^2$ intervals. Simulations using the Higher-Twist approach and the Generalized Higher-Twist approach are compared to the LO estimation of momentum broadening using $K=4$. Bands for eHIJING MC simulation represent statistical uncertainties. }
    \label{fig:EIC-DpT2xB}
\end{figure}

Finally, in Fig.~\ref{fig:EIC-DpT2xB}, we show the $p_T$-broadening of charged pions with $0.2<z_h<0.5$ as a function of $x_B$ for two different regions of $2<Q^2<6$ GeV$^2$.  Bands with crossed and dotted hatches are predictions from eHIJING with the GHT and HT approaches, respectively. The leading-order estimation, neglecting energy loss, is
\begin{align}
\Delta \langle p_T^2 \rangle \approx \frac{C_F}{C_A} Q_s^2(x_B, Q^2) \frac{\int_{0.2}^{0.5} z_h^2 \frac{d\sigma}{dz_h dx_B dQ^2}  dz_h}{ \int_{0.2}^{0.5} \frac{d\sigma}{dz_h dx_B dQ^2} dz_h }.
\end{align}
The qualitative increase of $\Delta \langle p_T^2 \rangle$ at small $x_B$ and large $Q^2$ from eHIJING simulations and the LO formula is similar. However, due to parton energy loss and radiative broadening, the differences are sizable. Therefore, understanding the radiative correction to transverse momentum broadening and reducing the theoretical uncertainty in the medium-induced radiation formula is important to interpret the data and extract $\hat{q}(x_B, Q^2)$ at future experiments.

\section{Summary and outlook}
\label{sec:summary}

Hadron and jet production in DIS with nuclear targets are key observables to understanding nuclear partonic structures in lepton-nucleus collisions. The gluon distribution at small Bj\"orken $x_B$ and its saturation phenomena are closely related to the measurable $p_T$ broadening of hadrons and jets. 
In the meantime, parton shower evolution and hadronization can also be modified by multiple parton scattering and induced gluon radiation. 
These effects affect the interpretation of the jet and hadron momentum broadening in terms of the gluon saturation phenomena.

In this paper, we develop the eHIJING Monte Carlo event generator to simulate the multiple-parton medium interaction and modified jet fragmentation process in electron-ion collisions. 
Jet-medium interactions are mediated by small-$x$ gluons whose distribution is modeled with a parametric transverse-momentum-dependent gluon distribution function $\phi_g(x_g, \bfk; Q_s)$. The saturation scale $Q_s$ in $\phi_g$ is determined self-consistently, which in turn fixes the jet transport parameter $\hat{q}_g = Q_s^2/L$ of the cold nuclear matter. The locations and the momentum transfer of the multiple collisions are sampled according to the differential collision rates.
With the exchange of a small-$x$ gluon, the nucleon that participates in the multiple scattering is assumed to be broken into a quark-diquark pair.

We then implement the soft gluon emission limit of the medium-modified parton splitting function, which are obtained in the higher-twist (HT) or the generalized higher-twist (GHT) approach. The medium-induced splitting with a transverse momentum larger than $Q_s$ is implemented in the Sudakov form factor of the $p_T$-ordered parton shower in Pythia8. Multiple medium-induced emissions with scales below $Q_s$ are sampled in a formation-time-ordered manner.
Finally, Lund string fragmentation is applied to the colorless system of the parton shower and the nuclear remnants from both the hard and multiple scatterings.

With a reasonable choice of parameters, the model can describe the collinear and transverse-momentum-dependent observables in SIDIS as measured by the CLAS, HERMES, and EMC experiments and their nuclear size dependence. The model can explain, in particular,  the interplay between elastic and radiative energy loss, and the transverse momentum broadening.
Furthermore, we demonstrate that it can also describe less inclusive observables, such as the nuclear modification of di-hadron production in $e$+$A$ collisions.
Of course, there are still discrepancies in the exclusive limit, for example, the $z_h\approx 1$ region of the single-hadron production or the $z_2\approx 0.5$ of the double-hadron production in nuclear collisions.

We also use eHIJING to study the $x_B$ and $Q^2$ dependence of the jet transport parameter through the single-hadron transverse momentum broadening. 
At small $x_B$ and low $Q^2$, the $p_T$ broadening agrees with LO analysis that directly connects to the saturation scale as probed by the leading quark. 
At high $Q^2$, the model also includes radiative effects on the $p_T$-broadening providing indirect access to the transport parameter.
The $p_T$ broadening of a single hadron is complicated by hadronization effects and is sensitive to the form of the fragmentation functions.
Therefore, one should also consider exploring the use of jet momentum broadening as an interesting area to apply eHIJING in the future. 

There are also several improvements to be considered. 
First, the strangeness and baryon production is not as satisfactory. 
This may demand several simultaneous improvements: medium-modified flavor production $g\rightarrow q\bar{q}$ and double scattering with a quark of the nucleus (flavor-changing processes), as well as the hadronic transport in low-energy collisions.
Second, another focus of the future EIC is the medium modification of heavy flavor production, and one should consider including mass effects in the multiple collisions and medium-modified parton radiation. 
The NLO hard cross-sections with shower are essential to describe the entire $p_T$-spectra at large $Q^2$ in the DIS, in particular at small $x_B$. This requires going beyond the LO process of photon-quark scattering and modified jet shower in the nuclear medium at top EIC energies.
Finally, we remark that this version of eHIJING focuses on the large-$x_B$ ($x_B\gg 0.1/A^{1/3}$) physics as required Eq. (\ref{condition-2}) that states the region of validity of the geometric picture used in eHIJING.
At small $x_B$ ($x_B\ll 0.01$), the power counting and the geometric picture change drastically, and there are event generators based on saturation physics developed for the small $x_B$ region. 
It would be ideal if the two approaches could be interpolated and fill the gap of event generation in the intermediate $x_B$ region in the future.

\acknowledgments

We thank Yayun He for the discussions.  This work is supported in part by the Director, Office of Energy Research, Office of High Energy and Nuclear Physics, Division of Nuclear Physics, of the U.S. Department of Energy under Contract No. DE-AC02-05CH11231 and No.  89233218CNA000001  and within the SURGE topical Collaboration, by the US National Science Foundation under Grant No. OAC-2004571 within the X-SCAPE Collaboration, by the National Science Foundation of China under Grant No. 12022512 and No. 12035007, by the Guangdong Major Project of Basic and Applied Basic Research No. 2020B0301030008. YYZ is supported by the CUHK-Shenzhen University development fund under Grant No. UDF01001859. WK is also supported by the US Department of Energy through the Office of Nuclear Physics and the LDRD program at Los Alamos National Laboratory. Los Alamos National Laboratory is operated by Triad National Security, LLC, for the National Nuclear Security Administration of the U.S. Department of Energy (Contract No. 89233218CNA000001). Computations are performed at the National Energy Research Scientific Computing Center (NERSC), a U.S. Department of Energy Office of Science User Facility operated under Contract No. DE-AC02- 05CH11231.
\newpage
\appendix
\section{The medium expectation value of the gluon field correlator}
The transverse-momentum distribution of the jet parton after a single gluon exchange with the medium can be obtained by computing the following probability (summed over final-state color, and averaged over spin)
\begin{widetext}
\begin{align}
 P &= \frac{1}{p^+}\left\langle\int \frac{d^4k}{(2\pi)^4} \frac{d^4q}{(2\pi)^4} g_s^2 \mathrm{Tr}(t^bt^a) \mathrm{Tr}\left\{ \frac{(-i)(\slashed{p}-\slashed{k})}{(p-k)^2-i\epsilon}\slashed{A}^{b*}(k)  \frac{\slashed{p}}{2} \slashed{A}^a(q) \frac{i(\slashed{p}-\slashed{q})}{(p-q)^2+i\epsilon} \frac{\slashed{n}}{2} \right\} \right\rangle_A \nonumber \\
&= \int dr^+ d^2\bfs \rho(r^+, \bfb+\bfs)\int \frac{dk^- d^2\bfk}{(2\pi)^3} \frac{dq^- d^2\bfq}{(2\pi)^3} g_s^2 \frac{C_F}{d_A}  \frac{e^{i(k-q)^-s^+} e^{-i(\bfk-\bfq)\cdot \bfs}}{[k^-+\frac{\bfk^2}{2p^+}+i\epsilon][q^-+\frac{\bfq^2}{2p^+}-i\epsilon]} \sum_a\langle N|(A^{-,a})^*(k) A^{-,a}(q) |N\rangle \nonumber \\
&\approx \int dr^+ \rho(r^+, \bfb)\int \frac{dk^- d^2\bfk}{(2\pi)^3} \frac{dq^-}{2\pi} g_s^2 \frac{C_F}{d_A}  \frac{e^{i(k-q)^-s^+}}{[k^-+\frac{\bfk^2}{2p^+}+i\epsilon][q^-+\frac{\bfk^2}{2p^+}-i\epsilon]} \sum_a\langle N|(A^{-,a})^*(k) A^{-,a}(q) |N\rangle  \nonumber \\
&= \int_0^\infty dr^+ \rho(r^+, \bfb)\int \frac{d^2\bfk}{(2\pi)^2}  g_s^2 \frac{C_F}{d_A} \sum_a\langle N|(A^{-,a})^*(x_g P_N^-, \bfk) A^{-,a}(x_g P_N^-, \bfk) |N\rangle  \nonumber \\
&\Rightarrow \int_0^\infty dr^+ \rho(r^+, \bfb)\int \frac{d^2\bfk}{(2\pi)^2}  g_s^2 \frac{C_F}{d_A} \frac{(2\pi)^2\phi_g(x_g, \bfk^2)}{4\pi\bfk^2}
\end{align}
\end{widetext}
when coupled to collinear partons highly boosted to the positive light-cone direction, the $k^+, q^+$ integration can be directly performed and set $x^-=0$ and $y^-=0$ in the $A$ field and only the $A^-$ component contributes at this power.
The gluon correlator is evaluated in a nuclear wave function $|A\rangle$. Due to color confinement, the correlator is only non-vanishing within the same nucleon $N$ at the impact parameter $\bfb$. The average over the nuclear wave function is reduced to that of a nucleon wave function $|N(s)\rangle$ with an integration over the propagation direction of the one-particle density of the nucleon $\rho(s)$. 
Assuming the density is a slowly varying function of $\bfs$, i.e., $\rho(s^+, \bfb+\bfs) = \rho(s^+,\bfb)+\bfs\cdot\nabla_\perp \rho(s^+, \bfb)$ on the scale of the nuclear size and note that $\bfk$ and $\bfq$ are of a momentum scale comparable or larger than the proton mass, then to the zeroth order of the gradient expansion the spatial integral of $\bfs$ imposes $\bfk=\bfq$.
Finally, the contour integration sets $k^-=q^-=\bfk^2/2p+=x_g p_N^-$, with $x_g = x_B\bfk^2/2p^+x_B p_N^-= x_B \bfk^2/Q^2$. It should be noted that in the presence of collinear radiation, $x_g$ can be different, but in the current eHIJING approximation, the same formula for $x_g$ is applied in both cases with or without collinear radiation.

\section{Comparison of the Pythia8 default versus the dipole recoil shower for DIS}
\label{sec:default-vs-dipole}
\begin{figure}
\centering
\includegraphics[width=\columnwidth]{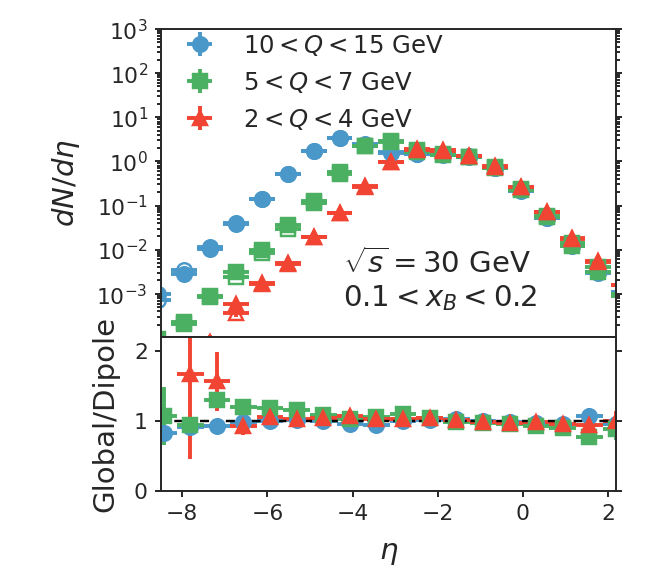}
    \caption{(Color online) The rapidity distribution in the lab frame. Comparison between Pythia8 simulations using the default shower and the dipole-recoil shower.}
    \label{fig:showers-dNdeta-lab}
\end{figure}

For a parton branching with given transverse momentum and momentum sharing fraction among the final-state partons, the one-to-two-body system $a\rightarrow b+c$ cannot fulfill the conversation of four momenta. In Pythia8, one solution to the problem is the ``dipole recoil'' method: for each radiating parton $a$, a recoiler parton $d$ is selected to form a dipole system. The process $a+d\rightarrow b+c+d$ can always restore energy-momentum conversation by properly shifting the four momenta of the recoiler $d$. For DIS, the IF (FI) dipole is formed by a radiator parton in the initial (final) state and a recoiler in the final (initial) state.
Another method to restore energy-momentum conservation is called the global recoil method, where the entire event is shifted accordingly after the branching $a\rightarrow b+c$.

The dipole recoil mode of Pythia8 is recommended for the study of DIS for two reasons. First, the dipole recoil approach does not change $Q^2=-(\ell_2-\ell_1)^2$ of the event. Second, the IF dipole alone can produce the singular structure of the NLO matrix element (ME). For this reason, the FI dipole emission is turned off in the dipole-recoil mode so that the first emission can be matched on the ME calculations.
However, in the current version of eHIJING, we choose to implement the medium modifications to the parton splitting function in the global recoil mode with both initial-state and final-state radiators. This is because the cold nuclear matter in DIS is a final-state effect and the radiator has to be the final-state partons. Of course, the drawbacks are that there may be double counting in certain phase space regions and that the global recoils obscure the precise determination of $Q^2$ of the event, especially when compared to data at relatively low $Q^2$. 

\begin{figure}
\centering
\includegraphics[width=\columnwidth]{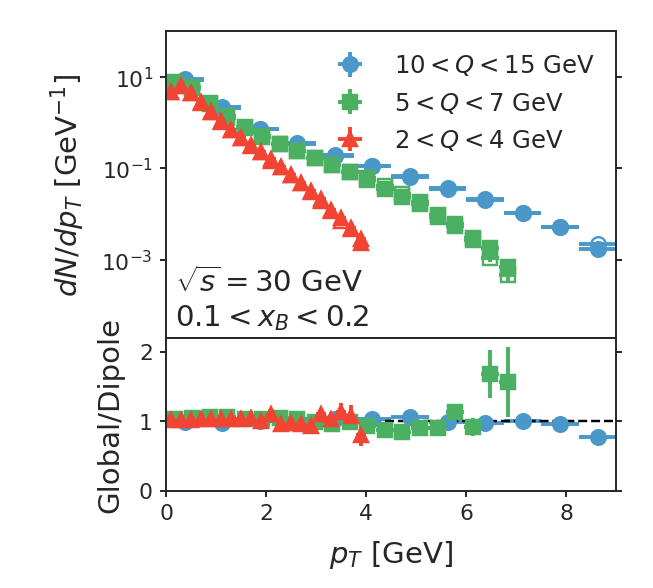}
    \caption{(Color online) The $p_T$ distribution in the lab frame. Comparison between Pythia8 simulations using the default shower and the dipole-recoil shower.}
    \label{fig:showers-dNdpT-lab}
\end{figure}

As a first attempt to include medium effects in DIS event generation, we will move forward with the global recoil option while keeping such problems in mind for further developments. Here, we investigate the difference between the two modes in the description of the single-hadron distribution in the SIDIS process of $e$+$p$ collisions. In Fig. \ref{fig:showers-dNdeta-lab}, we compare the rapidity distribution of the charged hadrons. The filled and open symbols are simulations with global recoil and dipole recoil, respectively. We tested three different $Q$ regions as shown in different colors and symbols. A similar comparison of the transverse momentum spectra in Fig. \ref{fig:showers-dNdpT-lab}. We find that the charged particle distribution in the lab frame is similar for the two recoil approaches.

\begin{figure}
    \centering
    \includegraphics[width=\columnwidth]{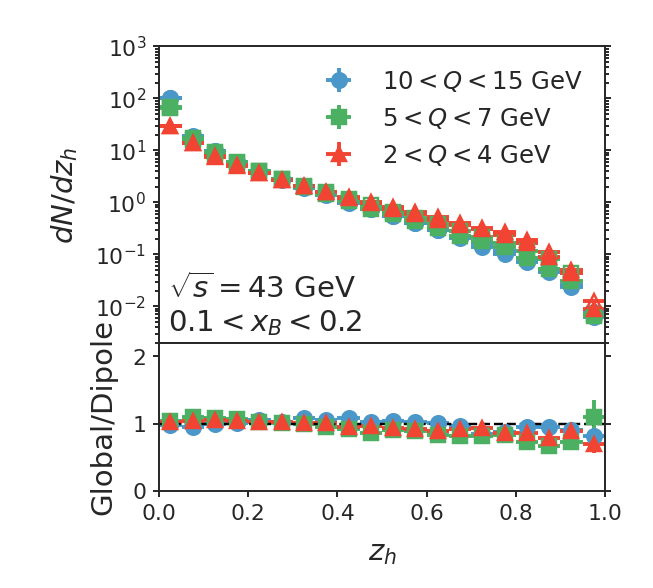}
    \caption{(Color online) Comparison of the $p_T$ spectra in the Breit frame from Pythia8 simulations using the default shower and the dipole-recoil shower.}
    \label{fig:showers-dNdz-Breit}
\end{figure}

\begin{figure}
    \centering
    \includegraphics[width=\columnwidth]{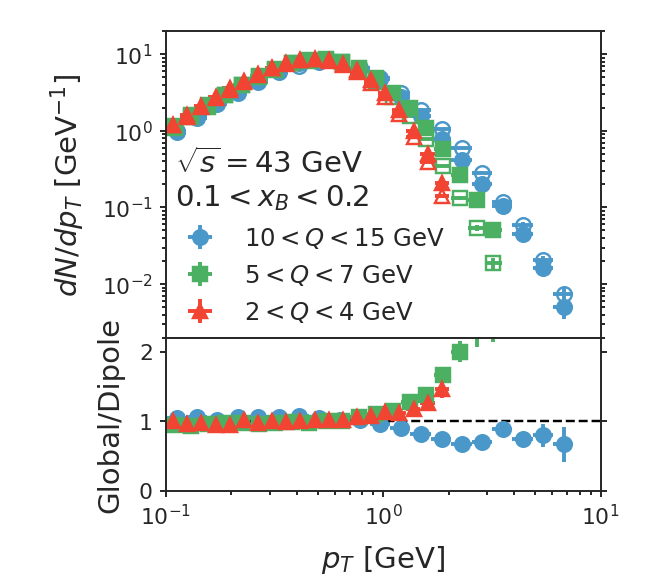}
    \caption{(Color online) Similar to Fig. \ref{fig:showers-dNdz-Breit} but for the $p_T$ distribution in the Breit frame.}
    \label{fig:showers-dNdpT-Breit}
\end{figure}

The situation is different if one presents the results in the Breit frame. Fig. \ref{fig:showers-dNdz-Breit} and \ref{fig:showers-dNdpT-Breit} shows the $z_h$ and $p_T$ distributions in the Breit frame.
There are notable differences in particle production close to the phase space boundary. The global recoil option produces fewer particles than the dipole recoil option when $z_h$ approaches unity. Nevertheless, the shape of $dN/dz$ between the two options is still similar and will not change the ratio $R_A$ too much.
For $dN/dp_T$ the difference is drastic when $p_T>1.5$ GeV at low $Q$ and the situation is improved for $10<Q<15$ GeV.
This is because that the hard $2\rightarrow 2$ scattering does not generate any transverse momentum in the Breit frame, making the $p_T$ spectra much more sensitive to the shower algorithm than the particle spectra in the laboratory frame.
Even though this should not affect our comparison to CLAS and HERMES data at low $p_T$, we do remind the users of the eHIJING generator that there are known issues in the baseline at large $p_T$ in the Breit frame.

\newpage
\bibliography{ref}

\end{document}